\documentclass{aastex631}

\usepackage{comment}

\usepackage{booktabs}

\shorttitle{Titan's middle atmosphere zonal winds}
\shortauthors{Marlin et al.}

\graphicspath{{./}{}}

\begin{document}

\title{Zonal winds in Titan's middle atmosphere from a stellar occultation observed with Keck adaptive optics}

\author[0009-0003-0670-5474]{Theresa C. Marlin}
\affiliation{California Institute of Technology \\
1200 E California Blvd \\
Pasadena, CA 91125, USA}

\author{Eliot F. Young}
\affiliation{Southwest Research Institute \\
1301 Walnut Street Suite 400 \\
Boulder, CO 80302, USA}

\author{Katherine de Kleer}
\affiliation{California Institute of Technology \\
1200 E California Blvd \\
Pasadena, CA 91125, USA}

\author{Martin Cordiner}
\affiliation{Catholic University of America}
\affiliation{NASA Goddard Space Flight Center}

\author[0000-0001-8621-6520]{Nicholas A. Lombardo}
\affiliation{Yale University}

\author{Imke de Pater}
\affiliation{University of California, 501 Campbell Hall, Berkeley CA 94720}

\author[0000-0001-9925-1050]{Juan M. Lora}
\affiliation{Yale University}

\author{Paul Corlies}
\affiliation{Spectral Sciences, Inc.}

\author{Richard Cosentino}
\affiliation{Space Telescope Science Institute}

\author{Conor Nixon}
\affiliation{Planetary Systems Laboratory, NASA Goddard Space Flight Center}

\author{Sébastien Rodriguez}
\affiliation{Université Paris Cité, Institut de Physique du Globe de Paris, CNRS, Paris, France}

\author[0000-0002-8178-1042]{Alexander Thelen}
\affiliation{California Institute of Technology \\
1200 E California Blvd \\
Pasadena, CA 91125, USA}

\begin{abstract}

We present spatially resolved Keck/NIRC2 images of a stellar occultation by Titan on September 5, 2022 and compare them to predictions from concurrent ALMA observations and a suite of General Circulation Model (GCM) simulations. ALMA data and GCM simulations can predict middle atmosphere zonal wind distributions, which in turn produce diagnostic occultation image sequences. We construct an occultation forward model using the temperature profile measured by the Huygens Atmospheric Structure Instrument, which is then distorted using latitudinal zonal wind profiles from the ALMA data or GCM simulations. The occultation forward model yields simulated light distributions around Titan’s limb, which we compare directly to the light distributions observed during the occultation. The GCM zonal wind profile corresponding to slightly before the time of the stellar occultation provides the best overall match to the data. The ALMA wind profile provides the best match to the occultation data when only the ingress data were considered, but is not the best match when data from ingress and egress are combined. Our data support the presence of stronger winds in the southern hemisphere during late northern summer.

\end{abstract}


\section{Introduction} \label{sec:intro}

Titan's thick, hazy atmosphere is the site of photochemical dissociation of the atmospheric constituents molecular nitrogen (N${_2}$) and methane (CH${_4}$), and subsequent recombination and synthesis of complex organic species \citep{willacy_vertical_2022, nixon_composition_2024}. Photochemical pathways in Titan's atmosphere generate heavier molecules which descend through the atmosphere, and eventually get deposited across Titan's geologically-varied surface \citep{brown_titan_2010,horst_titans_2017}. The intimate interactions of Titan's surface and atmosphere result in an active world \citep{mackenzie_titan_2021}, where molecules created by complex chemistry high in the atmosphere eventually reach the dunes, lakes, plains, and mountains of Titan, crusting the water-ice surface and settling in the polar lakes \citep{soderblom_correlations_2007,clark_detection_2010}.

The composition of Titan's atmosphere is also dependent on the satellite's atmospheric dynamics \citep{flasar_structure_2008}, which undergo seasonal variations over the course of Saturn's 29.5 Earth year orbit \citep{teanby_active_2012,teanby_titans_2008,teanby_vertical_2007}. We use solar longitude ($L_s$) as a metric of where Titan is in its orbit around the sun in order to compare data from across different times in Titan's year. $L_s = 0^{\circ}$ refers to northern vernal equinox, $L_s = 90^{\circ}$ is northern summer solstice, $L_s = 180^{\circ}$ is northern autumnal equinox and $L_s = 270^{\circ}$ is northern winter solstice. The middle atmosphere is thought to most often have a single thermally direct circulation cell that stretches from pole to pole, with upwelling at the summer pole and subsidence at the winter pole \citep{teanby_active_2012}. During vernal and autumnal equinox, the circulation cell reverses, and the atmosphere is predicted to briefly sustain two equator-to-pole circulation cells \citep{horst_titans_2017}. General Circulation Model (GCM) simulations likewise predict that the middle atmospheric zonal winds follow a cyclical pattern over the course of a Titan year, with rapid changes occurring around the equinoxes \citep{hourdin_numerical_1995,newman_stratospheric_2011,lombardo_influence_2023,lombardo_heat_2023}.

Complete characterization of Titan's atmospheric dynamics relies on accurate measurements of Titan's atmospheric winds, which are challenging to acquire directly. Efforts to study Titan's middle atmospheric zonal winds have included applying the thermal wind equation to \textit{Cassini} CIRS (Composite InfraRed Spectrometer) temperature measurements \citep{achterberg_titans_2008,achterberg_temporal_2011,achterberg2023temporal}, measurements of the Doppler shifts of atmospheric species' emission lines \citep[e.g.,][]{2019NatAs...3..614L, cordiner_detection_2020} and analysis of stellar occultation data \citep[e.g.,][]{sicardy_two_2006, zalucha_2003_2007}. Each of these methods has its shortcomings: CIRS is limited in temporal and spatial scope by \textit{Cassini}'s lifespan and Titan fly-bys and is also unable to retrieve wind speeds near the equator; stellar occultations are dependent on a fortuitous alignment of a sufficiently bright star, Titan, and an adequate telescope on a clear night; and Doppler shift measurements predict a single zonal wind profile for a range of altitudes that corresponds to the peak of the measured molecule's contribution function. However, when used in concert, these techniques provide insight into a challenging and essential region of Titan's atmosphere. In this paper, we present both spatially-resolved data from a 2022 stellar occultation ($L_s =160^{\circ}$) captured with adaptive optics and Doppler shift measurements of millimeter emission lines of the CH$_3$CN molecule, whose contribution function peaks around 300 km in Titan's atmosphere. We use a forward model to compare our occultation data to a wind field derived from the ALMA Doppler shift measurements, as well as to wind fields from an existing GCM simulation for Titan's middle atmosphere. The altitude ranges for these three methods are complementary, with occultation data probing altitudes of roughly 180-500 km, CH$_3$CN Doppler shift measurements centering around 200-400 km (with limited sensitivity up to 800 km), and GCM simulations providing predictions for altitudes up to 650 km.

\subsection{Titan's zonal winds via stellar occultations}
\label{stellar_occ_intro}
While occultations are often used to infer shape or ring properties of solar system bodies, a stellar occultation by a body with a thick atmosphere presents special opportunities for atmospheric characterization. During a stellar occultation, Titan's thick atmosphere refracts starlight rays, causing light ``spots" to be visible around the limb of the moon, even when an occulted star is physically located behind Titan's disk. The behavior and characteristics of these lightspots are sensitive to Titan's atmospheric density field between the stratosphere and mesosphere (altitudes of roughly 180--500 km): rays at lower altitudes are attenuated by differential refraction, while the density gradient at higher altitudes is too slight to sufficiently bend rays when the star is behind Titan's disk. Integrated line-of-sight refractivities can be calculated from densities, which allow tracing of starlight as it is bent by Titan's atmosphere. The location of the observed refracted starlight is highly sensitive to non-spherical atmospheric distortions, which are in turn related to atmospheric zonal wind patterns. Thus, by modeling the behavior of the starlight as it passes through Titan's atmosphere, Titan's zonal wind behavior can be estimated at the time of the occultation. 

The occultation shadow cast by Titan on Earth is not completely devoid of starlight: the light refracted by Titan’s limb fills the shadow in ways that depend on the non-spherical shape of Titan’s atmosphere. A uniformly rotating body with an atmosphere has an oblate shape that depends on the body’s gravitational field and the rotation period. A common approximation uses the gravitational $J_2$ term, the equatorial radius ($a$), the uniform angular velocity ($\omega$) and the surface gravity ($g$) to express the eccentricity ($e$) of the atmosphere \citep[Eq. 6 from][]{hubbard_occultation_1993}:

\begin{equation}
e = \frac{3}{2}J_2 + \frac{1}{2}\frac{a \omega^2}{g}.
\label{eccentricity}
\end{equation}

\begin{figure}[hbt!]
\centering
\includegraphics[width=10cm]{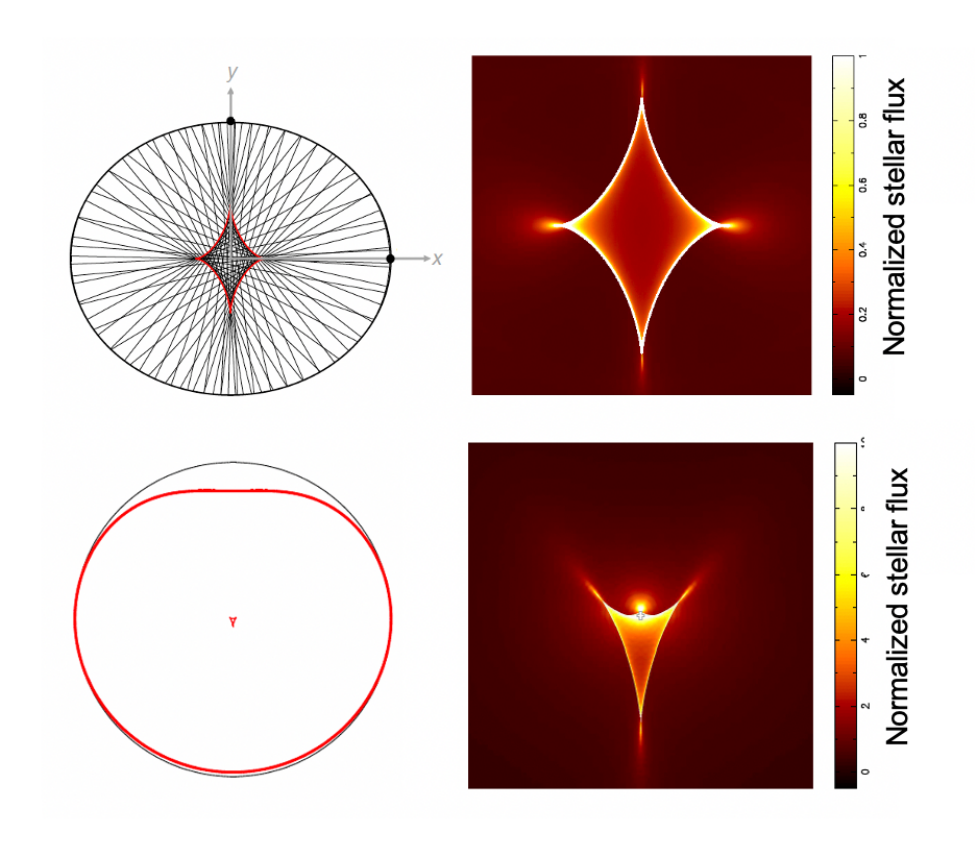}
\caption{(From Figs. 9 and 10 of \cite{sicardy_study_2022}). The top panels show the diamond-shaped caustic (concentration of light rays) that results from a planet with a uniformly rotating, oblate atmosphere (depicted as a black oval, top left). The black lines represent refracted starlight, and the caustic is highlighted in red. The top right depicts a magnified version of the caustic for the atmosphere on the left, showing the abrupt variations in flux that are present. Observation chords intersecting the caustic show rapid changes in flux at the center of an occultation event, allowing observers to determine the shape of the caustic. The bottom panels show the caustic that is formed in an alternative case: when the atmosphere is not uniformly rotating, but rather has a strong zonal wind in the north causing atmospheric distortion (highlighted by the red flattened circle outline). The distorted atmosphere produces a triangle-shaped caustic (the  red triangle in the middle of the bottom left image); the caustic and its flux variations are depicted in the bottom right. As in the case of the oblate atmosphere, observation chords that intersect the caustic capture variations in flux that allow mapping of the caustic. As caustics are highly sensitive to non-spherical distortions, observations of the central flash allow observers to determine the shape of Titan's atmosphere and therefore the zonal wind profile at the time of the occultation.).
\label{fig:sicardy_caustic}}
\end{figure}

Oblate planets have a bright diamond-shape caustic feature in the middle of the shadow (where ``caustic'' means a concentration of light rays), and observers on chords that intersect this caustic see bright features near the midpoint of the occultation, referred to as ``central flash'' features. The caustic and central flash features are highly sensitive to deviations from oblateness (Figure \ref{fig:sicardy_caustic}). The high sensitivity of central flash features to non-spherical atmospheric distortions and therefore zonal wind patterns allows zonal wind profiles to be studied via investigation of central flash features.

 Previous investigations \citep[e.g.,][]{hubbard_occultation_1993} have used central flash features to deduce zonal winds as a function of latitude. \citep{hubbard_occultation_1993} observed Titan occulting the bright ($V_{mag}$ 5.5) star 28 Sagittarius (28-Sgr), with the center of the occultation path crossing over densely populated areas with a number of observatories. Using a ``picket fence" series of observations located at different latitudes, observers were able to collect 23 lightcurves of Titan as it passed in front of 28-Sgr. Although no spatially-resolved data were obtained, \cite{hubbard_occultation_1993} acquired central flash data during the middle of the event from eight of their fifteen sites. They showed that these central flashes were inconsistent with a simple oblate model of Titan's atmosphere, regardless of whether Titan's solid-body rotation or a faster rotation (i.e., a super-rotating atmosphere) was assumed. Instead, their best-fit model required fast zonal winds with peak velocities of $\sim$175 m/s at latitudes near 65$^{\circ}$N and S, and much slower winds ($\sim$90 m/s) near the equator.

Adaptive Optics (AO) images of Titan during an occultation have only been obtained once before, when \cite{bouchez_seasonal_2004} observed Titan pass in front of a double star on 2001-12-20 ($L_s \sim 265^{\circ}$) with the PHARO instrument on the Hale telescope at Palomar Observatory. While 1-D lightcurves offer a single metric -- total flux as a function of time over the course of the occultation -- spatially-resolved occultation images reveal the distribution of light around the limb and the angle of offset between near and far limb spots. From these data, Bouchez derived a zonal wind profile that is qualitatively similar to that of \cite{hubbard_occultation_1993} (winds less than $\sim$80 m/s near the equator, winds over 180 m/s at latitudes that are 50° N and S of the minimum), but with different latitudes marking the maximum and minimum zonal winds \citep{bouchez_seasonal_2004}.

In this paper we report on image sequences taken during an occultation in which Titan’s disk was resolved with Keck's AO system. The 2022-09-05 observations at Maunakea are noteworthy in that the center of the occultation shadowpath passed within a few tens of km of the Keck II and Gemini North telescopes. The Titan-star offset in this occultation dataset is the smallest that has ever been captured with Adaptive Optics. 

\subsection{Titan's zonal winds via Doppler shift emission measurements}

The first direct detection of zonal winds in Titan's upper stratosphere/lower mesosphere ($z=300$--450~km) was by \cite{moreno_interferometric_2005}, using spatially resolved, ground-based microwave spectroscopy of Titan's entire Earth-facing hemisphere with the Plateau de Bure interferometer. The line-of-sight wind speeds at the limb were calculated from observed frequency shifts of molecular emission lines, which occur due to the Doppler effect. Observations conducted with the extended Submillimeter Array around northern spring equinox ($L_s=355^o$ and $L_s=5^o$) suggested a rapid increase in stratospheric wind speed following equinox, although the data were too coarse to be latitudinally resolved \citep{light2024measurements}. \cite{lellouch2019intense} used the Atacama Large Millimeter/submillimeter Array (ALMA) to derive Doppler wind maps for six gases, covering eight different altitudes in the range $z\sim300$--1000~km. The combined spectral/spatial resolution and sensitivity of ALMA allow the Doppler shifts of the observed emission lines to be measured as a function of position across Titan's disk, from which the zonal wind speeds as a function of latitude can be derived.  While the stratospheric ($\sim$ 50-400 km) zonal wind speeds were found to peak south of the equator, an unexpectedly intense, superrotating equatorial jet was found at thermospheric altitudes $\sim700$--1000 km. The time variability and dynamical stability of these jets was subsequently investigated in a followup ALMA study \citep{cordiner_detection_2020}. 

\subsection{Titan's zonal winds inferred from thermal wind measurements}

Most of what is assumed about Titan's middle atmospheric zonal winds is derived from the cyclostrophic thermal wind equation, which relates the meridional temperature gradient to the wind shear along an axis parallel to Titan's rotation axis \citep{flasar2005titan}.  Using this method relies on an accurate boundary condition for the zonal winds at the 10$^{3}$ Pa (0.01 bar) level in the atmosphere (100 km), which is assumed to be 4 $\times$ Titan's solid body rotation rate \citep{flasar2005titan,achterberg_titans_2008,achterberg_temporal_2011,sharkey2021potential,achterberg2023temporal}, consistent with measurements using the Huygens probe \citep{bird2005vertical}.

\cite{sharkey2021potential} used this method of inferring the zonal winds from thermal measurements to describe the seasonality of Titan's middle atmospheric jet from CIRS measurements. They found that the jet underwent substantial morphological changes over the duration of the \textit{Cassini} mission, with the core of the jet (the region with the strongest winds) migrating from the low winter latitudes (up to $\sim 30^{\circ}N$, with data unavailable near the equator) in 2007 to the high autumn latitudes (peak around $60^{\circ}S$) in 2015.  There was a similar change in the altitude of the jet, with the fastest winds found higher in the atmosphere near equinox ($\sim$500 km) than near solstice ($\sim$300 km).  The jet reached a maximum speed that exceeded 240 m s$^{-1}$ in mid autumn, and was weakest near solstice, with a speed closer to 180 m $s^{-1}$.

\subsection{Titan's zonal winds simulated with General Circulation Models}

GCMs are useful tools in assessing the roles that different physical processes play in driving the circulation of an atmosphere.  The first fully three-dimensional GCM of Titan's atmosphere was reported by \cite{hourdin_numerical_1995}. The model, while simple in its assumption of a meridionally uniform and static distribution of trace gases and aerosols, captured the large scale structure and seasonality of Titan's middle atmosphere.  The magnitude of the zonal winds, though, was substantially weaker than those inferred from \textit{Voyager 1} and \textit{Cassini} thermal measurements.  Likewise, this early model suffered from a relatively low model top (250 km), precluding an analysis of the full vertical extent of Titan's middle atmospheric jet.  TitanWRF, a three-dimensional GCM reported by \cite{newman_stratospheric_2011}, which utilized a higher model top and modified diffusion scheme, simulated faster zonal winds, though the simulated temperatures of the model exceeded those observed by Cassini by several tens of degrees Kelvin.  The impact of time-varying atmospheric opacity due to seasonal scale changes in Titan's atmospheric aerosol structure was studied by \cite{lebonnois2012titan}, which included a haze microphysical model coupled to the atmospheric dynamics.  Comparable to the simulations from \cite{newman_stratospheric_2011}, this model produced zonal winds with a magnitude comparable to that observed, but with an unrealistic middle atmospheric thermal structure.

The Titan Atmospheric Model \citep[TAM;][]{lora_gcm_2015} is a three-dimensional GCM that has recently been updated to improve the realism of the simulated middle atmosphere ($\sim$60-600 km) \citep{lombardo_influence_2023, lombardo_heat_2023}, including utilizing an observationally driven database of Titan's atmospheric composition and aerosol opacity (the Seasonally Varying Radiative Species Dataset [SVRS], \cite{lombardo_influence_2023}) to calculate realistic radiative heating rates, as well as parameterizations to enable a higher model top.  While the inclusion of the seasonality of the observed radiatively active components of the atmosphere increased the magnitude of the zonal winds by a few tens of m s$^{-1}$, the simulated jet core achieved a maximum speed of about 180 m s$^{-1}$, still short of the 240 m s$^{-1}$ inferred from thermal wind measurements in \cite{sharkey2021potential}.  The seasonal cycle of the zonal winds simulated by \cite{lombardo_heat_2023} closely resembles the seasonality inferred from thermal wind measurements; we make use of these simulations.

The zonal jet is typically strongest in the low latitudes (0-30$^o$) of the autumn or winter hemisphere \citep{shultis2022winter}. In late winter, the winds strengthen, and the jet core is found at higher latitudes and altitudes. Simultaneously, the low latitude winds also strengthen, leading to a substantial increase in the average wind speed throughout all of the middle atmosphere.  By equinox, the strongest winds are found in the spring mesospheric high latitudes, and the jet core doesn't migrate to the opposite hemisphere until well into that hemisphere's autumn \citep{lombardo_influence_2023}. By the middle of autumn, the jet core is descending and weakening, leading to the observed mid winter weakened state of the middle atmospheric jet \citep{shultis2022winter}.

\section{Data collection, reduction, and processing}

\subsection{Adaptive optics imaging of the stellar occultation}

\begin{figure}[hbt!]
\centering
\includegraphics[width=10cm]{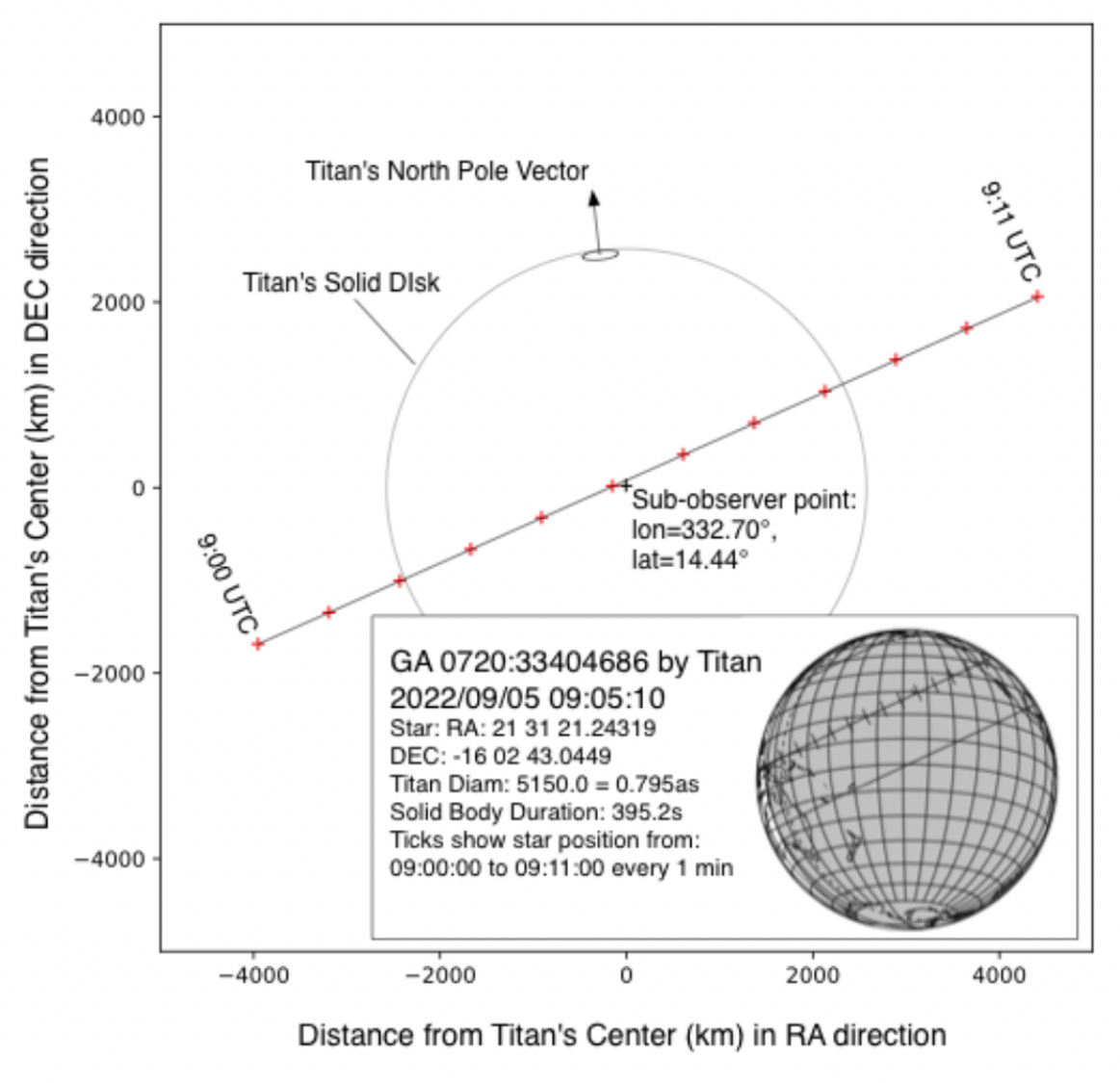}
\caption{Main figure: Path of the star (from the vantage point of Maunakea) as it is occulted by Titan. Red crosses represent one minute increments. The center of the occultation shadow path passes within fifty kilometers of Maunakea, making this site ideal for perceiving the ``central flash" (rapidly changing stellar refraction patterns) that occurs near the center of the occultation shadow path. Inset: full shadow path of star 2MASS21312124-1602427, K$_{mag}$ = 12.9. Hatched line represents the center of the shadow path; solid lines are the outer edges. \label{fig:occ_schem}}
\end{figure}

On September 5, 2022, Titan occulted a 12.9 K$_{mag}$ star (Figure \ref{fig:occ_schem}). We collected data from 7:51 to 10:21 UTC (second quarter-night) on September 5, 2022 using the NIRC2 instrument on Keck II. The telescope tracked Titan and the AO correction used Titan itself as the guide star. From 08:37:32 to 09:20:01, 245 frames of Titan were collected, each with an exposure time of four seconds and an average cadence of one frame per nine seconds. Data were collected in the Kp filter (1.90-2.35 $\mu$m) with frame size of 512 $\times$ 512 pixels. The frames spanned the entirety of the occultation (09:02-09:08 UT), as well as a span of time surrounding the occultation where the occulted star was visible beside Titan. Twenty sky frames (each also with an exposure time of 4 seconds) were collected, half before and half after the occultation sequence. After the occulted star disappeared from the field of view, we captured narrowband images of Titan (in the Br-gamma, H2 1-0, and He-IB filters) with integration times of 60 seconds, 60 seconds, and 30 seconds, respectively. These wavelength regions probe different regions of Titan’s atmosphere and allow monitoring of any transient haze activity occurring during the night of the occultation. We additionally took images of photometric calibration stars HST813-D for the Kp filter (integration time = 0.8 seconds, co-adds = 10) and HD201941 for the narrowband filters (integration time = 0.181 seconds, coadds = 10). 

Spatially-resolved occultation data were concurrently collected with the NIRI instrument on the Gemini-North telescope, also at a cadence of roughly one frame per nine seconds. However, the Gemini images have time stamps that are only accurate to 0.5 seconds (compared to tens of microseconds for the Keck time stamps). Given the event velocity of 13.8 km/sec, the distribution of light around Titan’s limb changes significantly in 0.5 seconds. We plan to evaluate whether separate file creation time stamps from the operating system can improve the Gemini timing.

\subsection{Occultation data reduction and processing}\label{sec:process}

We reduced Keck/NIRC2 data using the \texttt{nirc2reduce} Python package \citep{molter_emolternirc2_reduce_2024}. Reduction steps included median sky subtraction, application of a flat field, construction of a bad pixel map and subsequent bad pixel removal, cosmic ray removal, and flux calibration using the photometric standard stars.

\begin{figure}[hbt!]
\centering
\includegraphics[width=10cm]{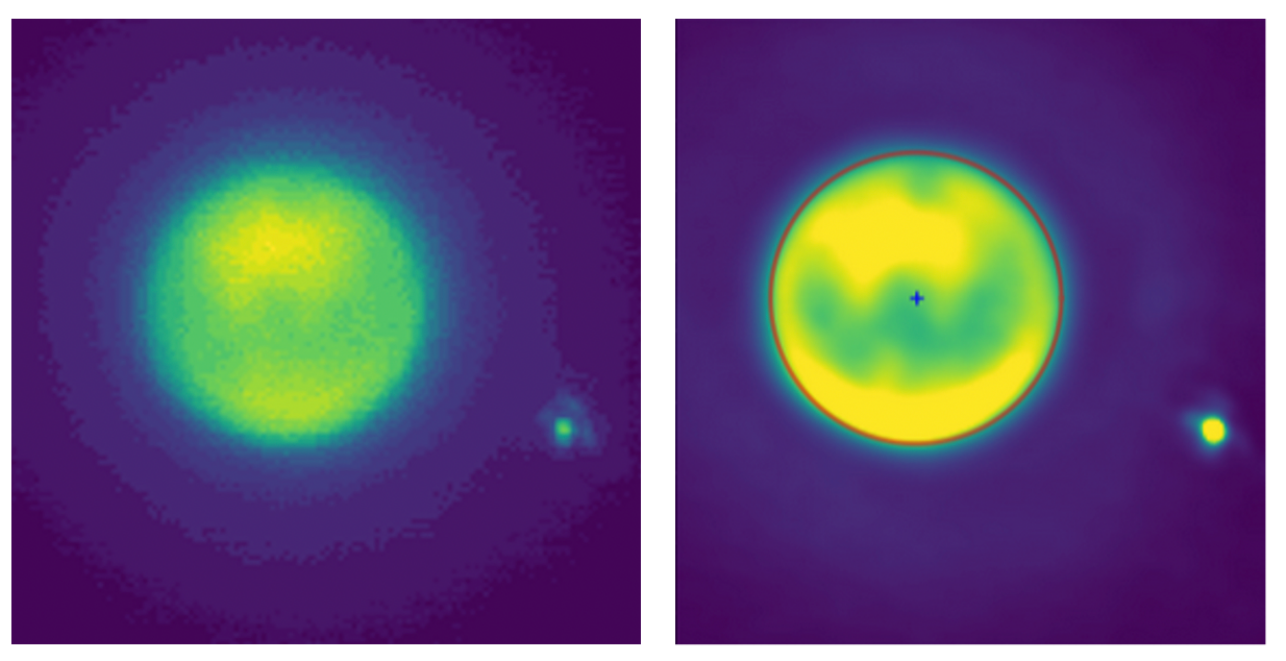}
\caption{Left: Reduced data from pre-event. Scattered light around both Titan and the star are caused by Earth's atmosphere. Right: The same frame following Richardson-Lucy deconvolution. Circle center and radius determined by fitting a circle to radial gradient minima around the limb. \label{fig:pre_post_deconv}}
\end{figure}

In order to better isolate and characterize refracted starlight around Titan’s disk, we implemented additional, custom processing steps. We used Richardson-Lucy image deconvolution \citep{Richardson_deconv,lucy_deconv} to sharpen the data and reduce the effects of Earth’s atmosphere on the observed images (Figure \ref{fig:pre_post_deconv}). Deconvolution first required defining a point-spread function (PSF). To find the best PSF for deconvolution, we studied the effectiveness of different PSFs in frames before and after occultation, when the occulted star was in the frame alongside Titan. The PSFs considered were: a frame-to-frame PSF of the star (where the PSF varied by frame), a universal PSF where the sharpest star (as determined by the lowest width of distributed light) was used as PSF for all frames, a frame-to-frame Gaussian, and a universal Gaussian. We ultimately used a universal PSF derived from the frame with the sharpest star, as this gave the sharpest deconvolved disks. For image deconvolution, we used the richardson\_lucy function from the skimage.restoration image restoration module by Scikit-Image \citep{van_der_walt_scikit-image_2014}. 

Following deconvolution, we stacked all images to minimize the slight drift of Titan across the field of view during observations. Stacking was accomplished by calculating the gradient along radial lines from the center for each frame. Then, a circle was fit to the minimum radial gradient values for all angles around the circle (the minimum -- or most negative -- radial gradient value corresponds to the sharpest change in flux, a proxy for the edge of the limb). Using the shift2d function from the fft\_tools.shift module of the image\_registration package by PyPI \citep{ginsburg_image-registration_2023}, the images were then aligned so that the (x,y) coordinates of the center of the fitted circles were stacked on top of each other. 

\begin{figure}[hbt]
\centering
\includegraphics[width=10cm]{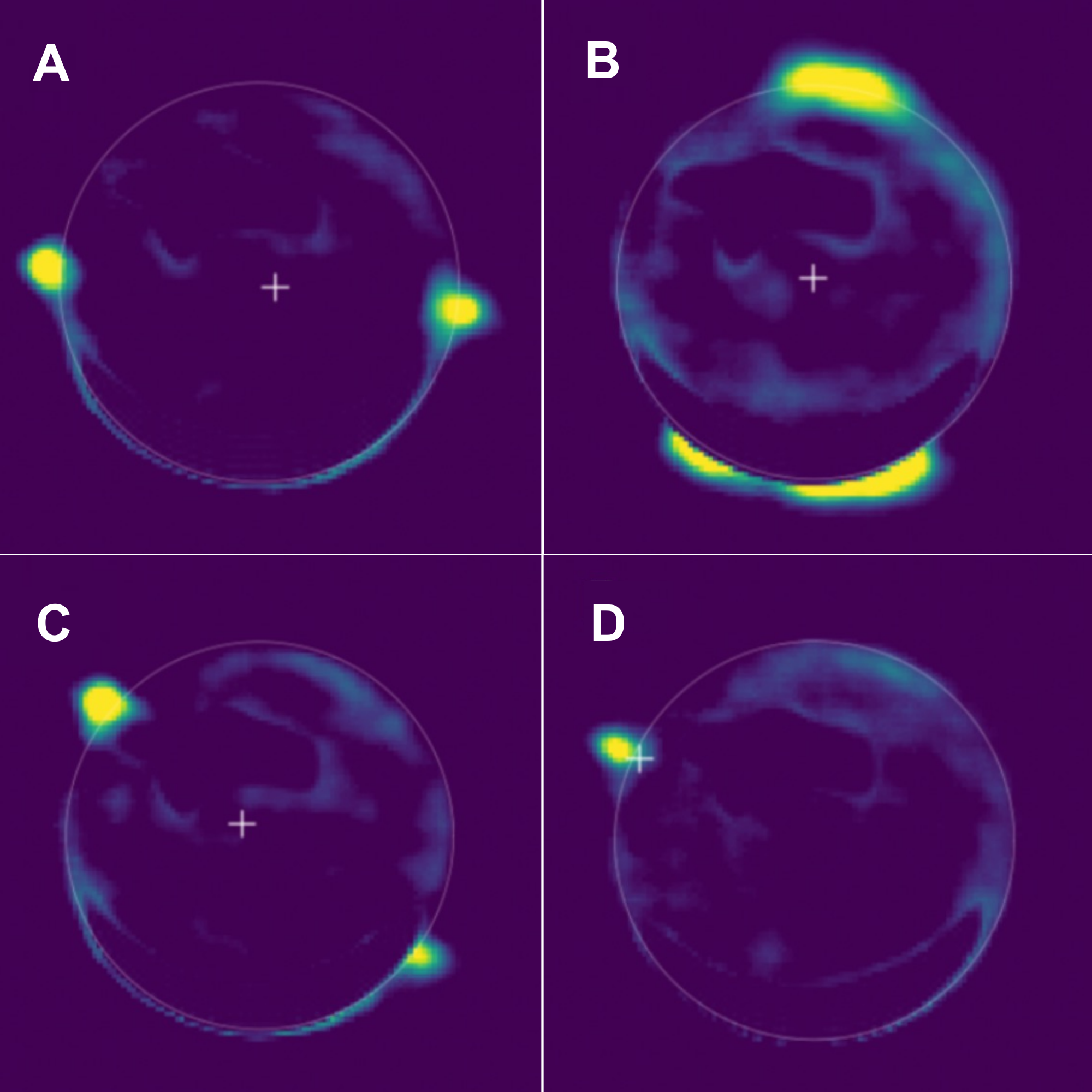}
\caption{Processed (deconvolved, disk-subtracted) data from the occultation. Panels A-C are from near the midpoint of the occultation, with Panel B being the most central frame of the occultation that we captured. In the middle of the event, multiple refracted lightspots are visible which change position rapidly with the cadence of the images. Panel D is from shortly before egress. The position of the refracted lightspot is aligned with the chord of the occultation.}
\label{animated_occ}
\end{figure}

To isolate refracted starlight during the occultation, it is necessary to subtract the dominant light coming from Titan's disk in each frame. To achieve this, we determined the median disk emission (calculated from the reduced, deconvolved, and stacked frames) and subtracted it from all of the frames so that only refracted starlight along the limb was visible in each frame. Individual frames from the occultation (post-processing) are shown in Figure \ref{animated_occ}.

\begin{figure}[hbt!]
\centering
\includegraphics[width=15cm]{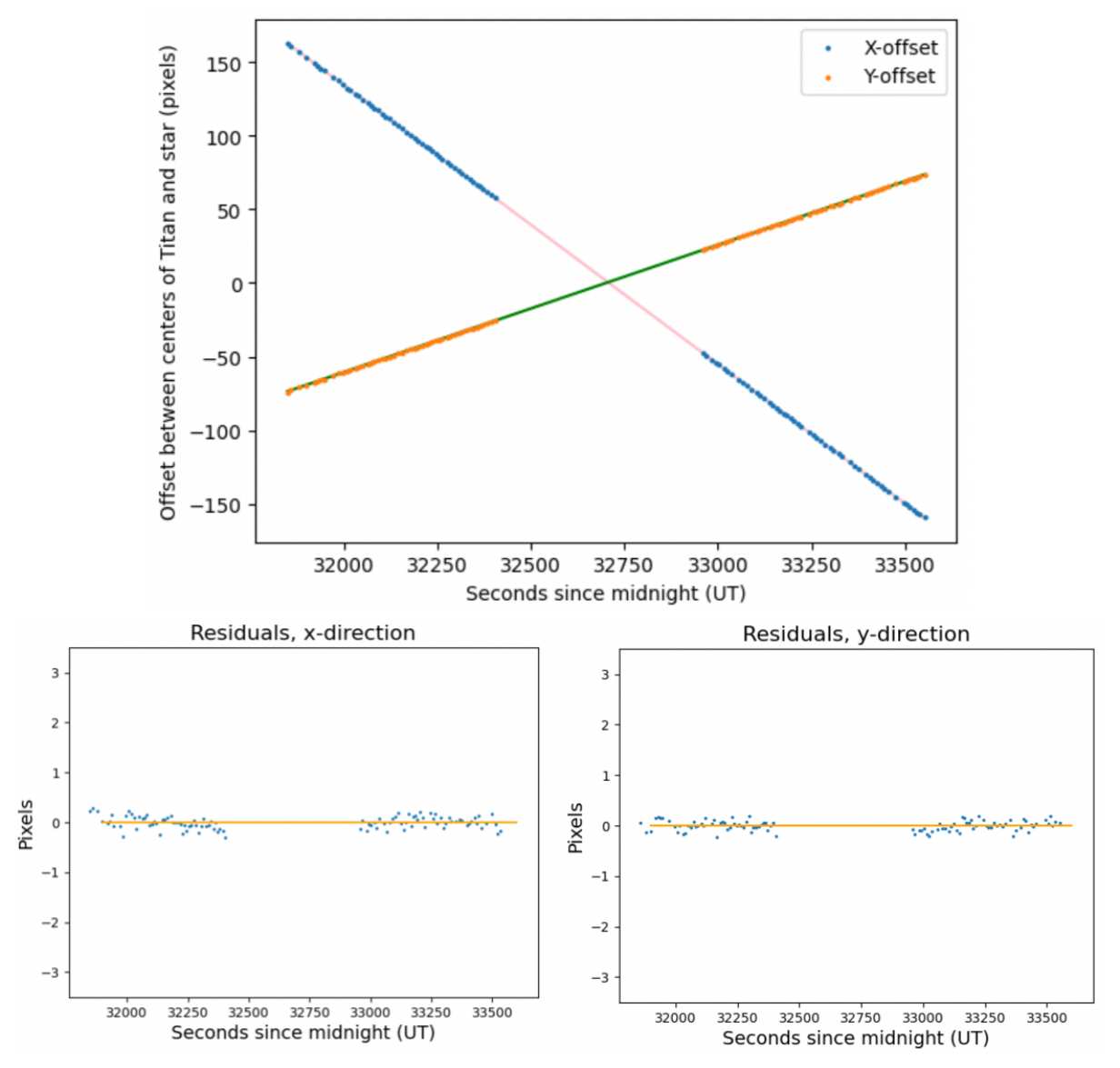}
\caption{Top: Linear fits (green, pink lines) to calculate x- and y-offsets between the centers of Titan and the occulted star. Bottom: Residuals in the x- (left) and y-directions (right) following the subtraction of the fit.\label{fig:star_titan_pos}}
\end{figure}

The forward modeling effort (see Section \ref{forward_model}) required the extraction of an additional parameter from the data: the distance in the detector plane between the center of the star and the center of Titan's disk. From the deconvolved frames before and after the occultation, the star was isolated in a 9$\times$9 pixel subframe (x-direction) or an 11$\times$11 pixel subframe (y-direction). These star subframe widths were selected based on which led to the lowest variability in centroid fitting. For calculating the centroid of the star in each frame, we used several different methods of centroid fitting, including two one-dimensional Gaussians, a two-dimensional Gaussian, center of mass, and a quadratic polynomial. Two one-dimensional Gaussians led to centroids with the least variability. The centers of Titan in each frame were calculated as the center coordinates of the circle fitted to the minimized radial gradients in each frame described above. Titan-star offsets were calculated for all pre- and post-occultation frames by taking the difference of the star centroid and the fitted circle center for Titan's disk. Linear fits were applied to both the x- and y- directions (as a function of time) in order to interpolate the x- and y-offsets during the frames when the star was not visible (i.e., being actively occulted by Titan) (Figure \ref{fig:star_titan_pos}). The $\Delta$x and $\Delta$y offsets were then converted into $\Delta$RA and $\Delta$DEC offsets for occultation frames by using the plate scale of the observations and the distance between Titan and Earth (taken from JPL HORIZONS). 

\subsection{ALMA observations and data reduction}
\label{sec:alma}

ALMA observations of Titan were obtained on UT 2022-09-06 between 02:29--05:48 (less than 24 hours after the occultation observation) using the Band 9 (602--720~GHz; $\sim0.42$--0.5~mm) receivers, covering 4 lines of the CH$_3$CN $J=35-34$ rotational band (between 643.1--643.3~GHz) and 4 lines of the $J=36-35$ band (between 661.4--661.7~GHz). The correlator was set to a spectral resolution of 122 kHz, corresponding to $\Delta{v}=55$--57~m\,s$^{-1}$. Weather conditions were very good, with a zenith precipitable water vapor of 0.22--0.32~mm.

The interferometric data were flagged and calibrated in the Common Astronomy Software Applications (CASA) package \citep{casa_2008} using standard scripts supplied by the Joint ALMA Observatory. Additional self-calibration of the phases was performed with respect to the Butler-JPL-Horizons 2012 Titan model included with CASA$\footnote{https://science.nrao.edu/facilities/alma/aboutALMA/Technology/ALMA\_Memo\_Series/alma594/memo594.pdf}$, using separate continuum windows adjacent to the main spectral windows of interest (centered at 649~GHz and 666~GHz, respectively).

\begin{figure}[h!]
\centering
\includegraphics[width=6.5cm]{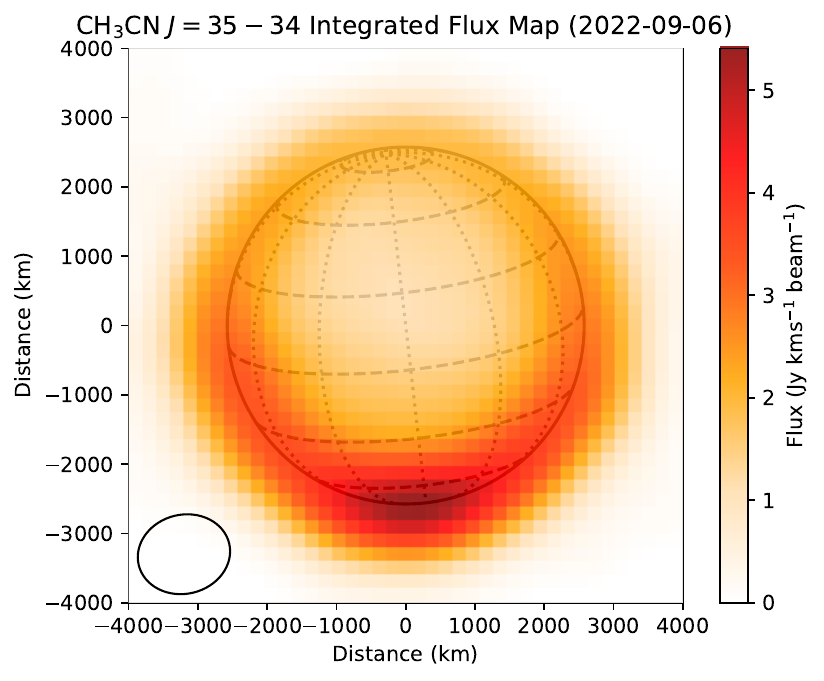}
\includegraphics[width=8cm]{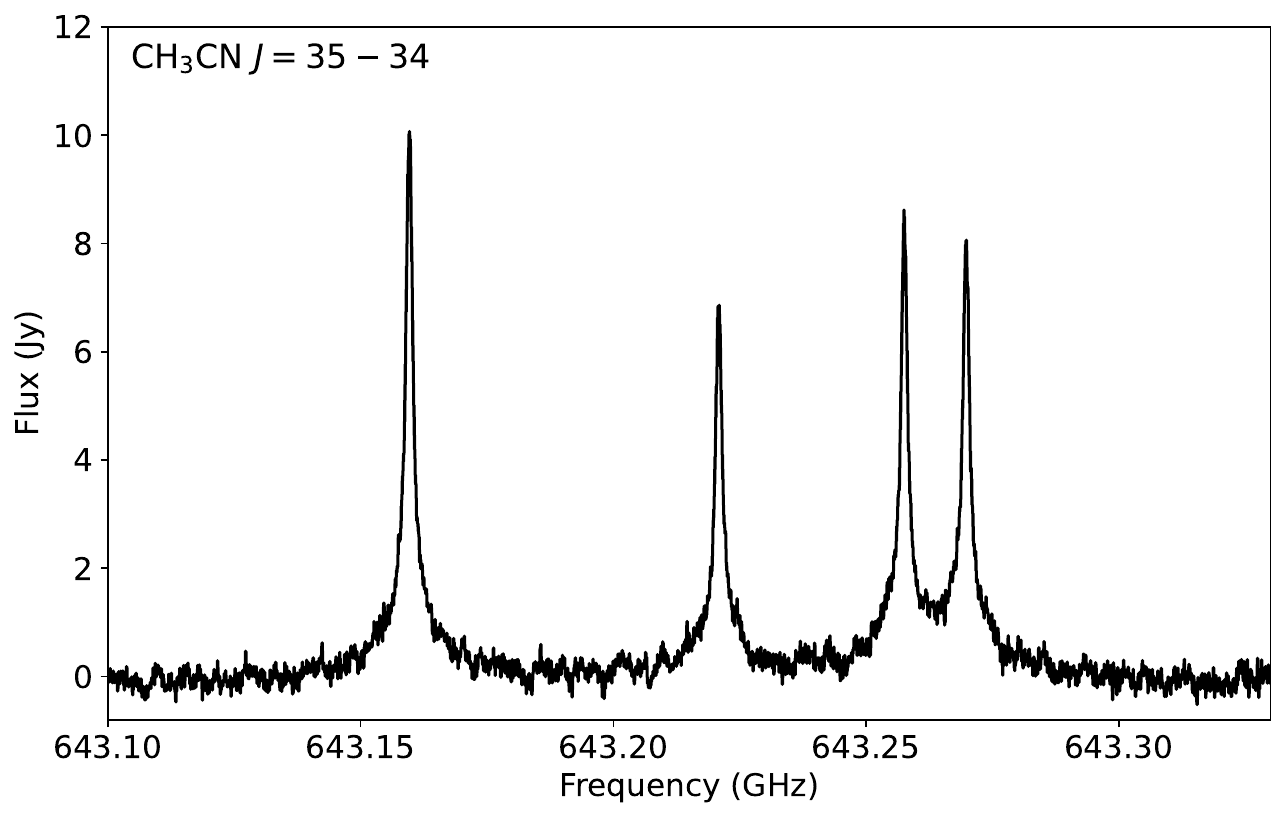}
\includegraphics[width=6.5cm]{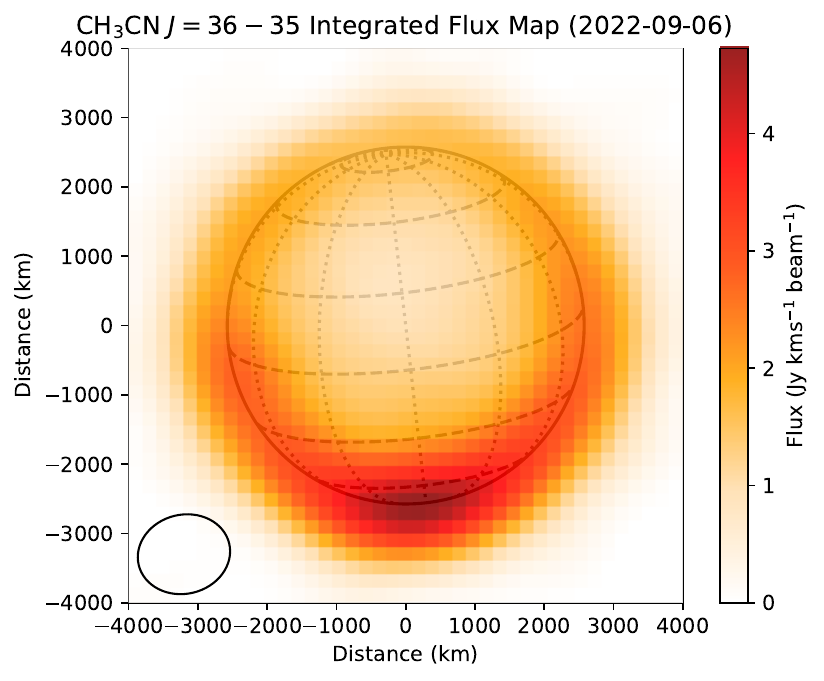}
\includegraphics[width=8cm]{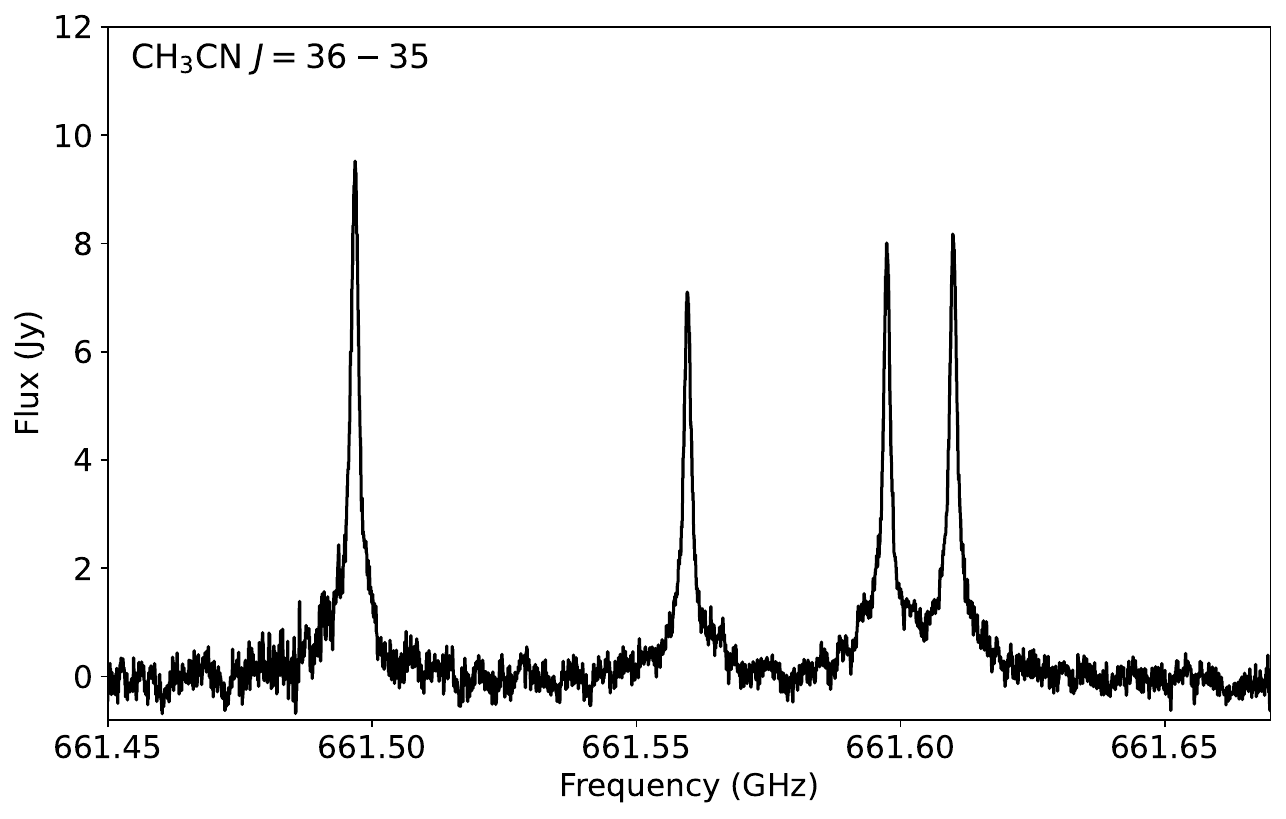}

\caption{Continuum-subtracted ALMA observations of Titan on 2022-09-06, showing the spectrally integrated line flux maps for the CH$_3$CN $J=35-34$ and $J=36-35$ bands (left panels) and corresponding spectra (right panels), integrated within a $1.2''$-diameter circular aperture centered on Titan. The FWHM of the spatial resolution element (ellipse) is shown in the lower left of the maps.\label{fig:almadata}}
\end{figure}

Prior to imaging, the Titan continuum was subtracted from the interferometric visibility data using a 2nd order polynomial fit to the line-free (pseudo-)continuum regions in each spectral window. The spectral image data were deconvolved (per channel) using the CASA {\tt tclean} Hogb{\"o}m algorithm with Briggs weighting (robust = 0.5) and a threshold twice the RMS noise ($\sigma=29$~mJy at 643 GHz and 38~mJy at 662~GHz). Image restoration was performed for both CH$_3$CN bands using the smallest common elliptical Gaussian beam, of FWHM $0.21'' \times 0.18''$. For plotting purposes, the data were projected onto a spatial grid in the plane of the sky, according to Titan's geocentric distance of 8.94~au. The spectrally integrated line flux maps and disk-integrated spectra for the two CH$_3$CN rotational bands are shown in Figure \ref{fig:almadata}.

\subsection{ALMA wind retrieval}

We followed a similar strategy to \cite{cordiner_detection_2020} to retrieve the latitudinal wind velocity profile based on fits to the observed spectral lines to determine their Doppler shifts. The Gaussian line fitting method was updated to use the Moffat function \citep{moffat1969theoretical} instead, which was found by \cite{light2024measurements} to provide a significantly better fit to the CH$_3$CN lineshapes than Gaussian or Voigt functions, especially in the presence of pressure-broadened Lorentzian line wings. We found that the Moffat function was able to fully reproduce our observed line profiles (within the noise), resulting in a more accurate determination of the line Doppler shift than the conventional Gaussian fitting method.

All eight CH$_3$CN spectral lines were fitted simultaneously for all pixels in the ALMA images, using a Levenberg-Marquardt least-squares minimization routine written in Python, based on the MPFIT package \citep{markwardt2009non}. All parameters of the Moffat function were allowed to vary independently for each line, apart from the velocity offset parameter, for which a common (variable) value was used for all lines. The uncertainty on the Doppler shift for each pixel was derived from the diagonal elements of the covariance matrix.

A significant ($\approx+30$ m\,s$^{-1}$) offset in the spectral line Doppler shifts was identified at the center of Titan's disk (and along the direction of its polar axis).  This is attributed to an error in the cable routing of the ALMA rubidium maser at the time of our observations, which is likely to have introduced an erroneous and unknown frequency shift in our observations, accompanied by a slight broadening of the spectral lines by a few tens of meters per second (R. Loomis, private communication).  The magnitude of the shift was calculated by taking the average of the fitted line-of-sight velocities inside Titan's limb. This ``zero point shift'' was then subtracted from the individual pixel Doppler shifts for the subsequent data analysis steps. No significant evidence for latitudinal asymmetry in Titan's zonal winds was reported by the detailed studies of \cite{lellouch2019intense}  and \cite{cordiner_detection_2020}. Our ALMA data analysis here therefore also inherently assumes that the zonal winds are symmetric with respect to Titan's polar axis. 

To determine the altitudinal sensitivity of our ALMA zonal wind measurements, we used the Non-linear optimal Estimator for MultivariatE spectral analySIS (NEMESIS) to generate a detailed radiative transfer model for the observed CH$_3$CN emission \citep{2008JQSRT.109.1136I}. We followed a similar modeling strategy to \cite{thelen2019abundance}, updated to incorporate the more recent CH$_3$CN abundance profile from \cite{nosowitz_improved_2025} for the \textit{a priori} atmospheric state. 

\begin{figure}[h!]
\centering
\includegraphics[width=9cm]{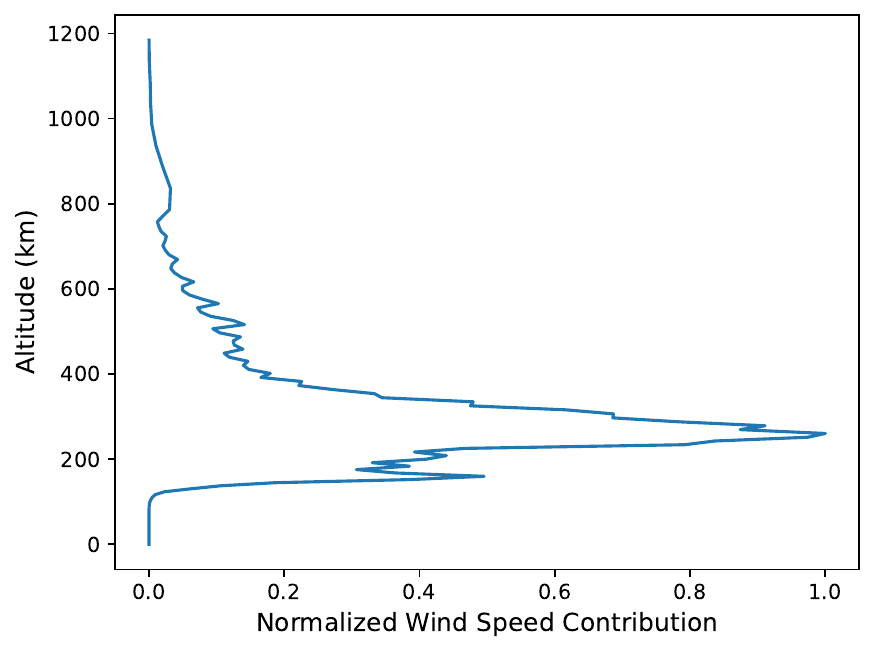}
\caption{Peak-normalized wind contribution function for the observed CH$_3$CN $J=35-34$ transitions at 643 GHz. \label{fig:contrib}}
\end{figure}

The calibrated ALMA data cubes were re-imaged prior to continuum subtraction, in the same way as described in Section \ref{sec:alma}. Disk-averaged spectra were then extracted using a circular aperture with a diameter of 1.5$''$, which includes flux from up to the top of Titan's atmosphere plus twice the standard deviation of the restoring beam's long axis, to account for all the significant emission. Titan's disk (plus atmosphere) was divided into 35 annuli corresponding to emission angles between 0-75 degrees, covering the center of Titan up to a tangential altitude of 1050 km (the vertical extent of our model was 1200 km, but emission is negligible above 1000 km). The atmospheric temperature profile was obtained from \cite{2020ApJ...903L..22T}. The observed ALMA data (including the continuum) were scaled by a constant factor of 0.982 to produce a match between the modeled and observed continuum levels. This scaling mitigates errors on the ALMA flux scale (which may be as much as $\approx10$\%, due to uncertainty in the amplitude calibrator flux), as well as accounting for small errors in the assumed (\textit{a priori}) temperature around the tropopause, which determines the model continuum level. The CH$_3$CN abundance profile was continuously retrieved at each altitude to obtain a fit to the observed $J=35-34$ spectral line profiles. This abundance profile formed the basis of our ``wind contribution function'' calculation, which was obtained from the corresponding NEMESIS functional derivatives ($dQ/dz$) of flux with respect to altitude, as a function of frequency. The $dQ/dz$ values were convolved with the gradient of the (continuum-subtracted) model spectral line profile, to provide weighting according to the frequencies contributing most to the retrieved line Doppler shifts (see \emph{e.g.} \cite{lellouch2019intense}), which were then averaged and peak-normalized.  The resulting, weighted wind contribution function is shown in Figure \ref{fig:contrib}.  The corresponding average emission altitude is 303 km, with sensitivity to winds primarily in the range $\sim150$-450~km. Due to the similarity of the line opacities and energy levels, this range is also applicable to the $J=36-35$ band.

Due to the effects of beam-smearing, combined with Titan's axial tilt, extraction of the intrinsic zonal wind field requires deconvolution of the Doppler shift data. We therefore followed a similar method to \cite{cordiner_detection_2020} to model the Doppler map and extract a latitudinal wind speed profile using a simplified 3D radiative transfer model. The {\tt atmWindMapGaussFit} code of \cite{cordiner_detection_2020} was modified to allow a continuously variable wind speed as a function of latitude ($v(l)$), instead of the Gaussian profile previously assumed by those authors. The $v(l)$ profile was generated by joining a set of $i$ points ($v_i(l_i)$, equally spaced in $l$), by a segmented linear function. This function was then smoothed by a Gaussian function of FWHM equal to the spacing between the points. An exponential taper was applied at the ends of the $v(l)$ profile (with an exponential smoothing scale length of 1$^{\circ}$), to ensure a velocity of zero at the poles. Each velocity point $v_i$ was then optimized using MPFIT, to obtain the least-squares solution and associated error estimates.

To maximize the information content in the $v(l)$ retrieval, we experimented in fitting the observations with profiles incorporating different numbers of variable latitudinal points $i$. Seven latitudinal points (with $\Delta{(l)}=30^{\circ}$ separation) were found to produce a good fit to the observed Doppler map, while still allowing the retrieved profile to be well constrained by the data (\emph{i.e.} keeping the resulting $v(l)$ errors reasonably small with respect to variations in the retrieved profile). At the equator, this $\Delta({l})$ interval corresponds to a distance of $0.2''$ on the sky, which is similar to the spatial resolution of our ALMA data.

\section{Forward modeling}

\subsection{Forward modeling of a stellar occultation}
\label{forward_model}

\cite{sicardy_study_2022} presents a comprehensive overview of occultations (including Titan as a detailed example). Among other things, this review paper shows how central caustics (concentrations of light rays, see Section \ref{stellar_occ_intro}) form in the middle of the shadow. The bending angles of the starlight being refracted around Titan are perpendicular to the contours of constant gas density around the limb. This means that small changes to the shape of the atmosphere lead to large changes in the shape of the central caustic and (in cases where the occulting object is resolved) the distribution of refracted light around the limb. During most of the occultation the refracted light is grouped into two spots, called the ``near limb'' and the ``far limb,'' referring to the limbs closer to and farther from the star, respectively. These spots appear wherever the sky-plane projection of the lines from the star to the limb are perpendicular to the contours of constant density (isopycnals). The locations of the near- and far-limb spots change as Titan moves across the star, sometimes rapidly near mid-event. 

Section \ref{zw_to_dist} describes how zonal winds change the isobars as a function of latitude. We use a quantity called $\Delta r$, the level surface radius, to describe distortions from a spherical atmosphere. On Titan, the changes to the level surface ($\Delta r$) amount to tens of km -- small compared to Titan's radius, but since it is the \emph{slopes} of the level surface that control where the near- and far-limb spots appear, the shapes of the isobars on Titan affect the spot locations in easily observable ways. The image sequence taken during the occultation shows the progression of light around Titan’s limb during an occultation and lets us compare those distributions to modeled images based on ALMA- or GCM-based zonal wind distributions. Although isobars and isopycnals are not identical, applying the isobar distortions to the bending angle fields over small temperature ranges gives undetectable errors.

Our forward model is constructed based on ``ray tracing'' -- constructing 2-D screens of Titan’s atmosphere (as would be observed by the telescope), and tracing rays of light as they pass through the screens.  We use the process outlined below, beginning with models of Titan’s atmosphere and ending up with bending angles as functions of altitude and latitude. To speed up this calculation, we generate a 1-D function for bending angle vs. radius. We accommodate the anisotropies due to zonal winds by distorting the radius field as a function of latitude, then calculating a 3-D field of bending angles around Titan's limb.

\begin{figure}[hbt!]
\centering
\includegraphics[width=18cm]{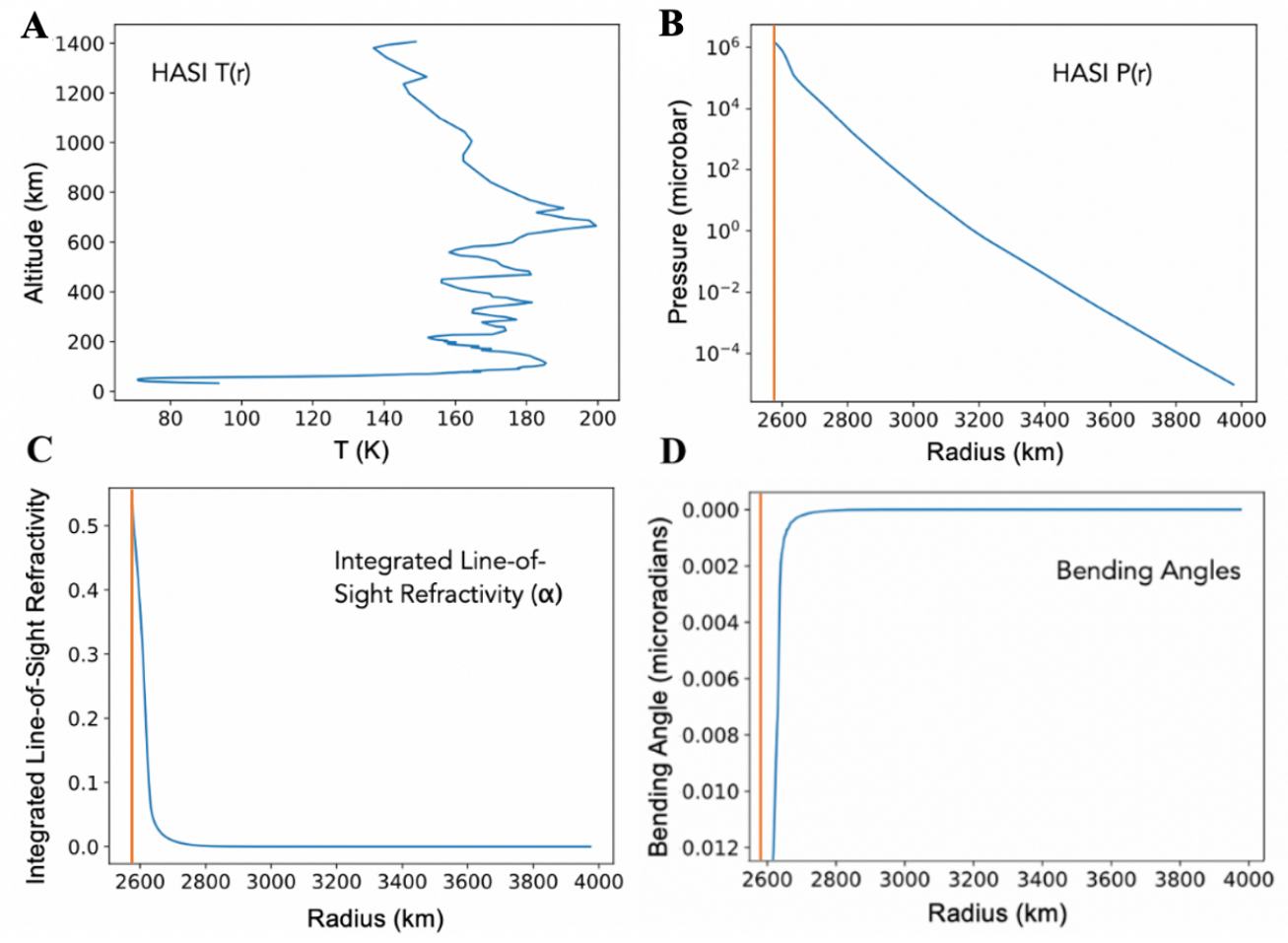}
\caption{A: The atmospheric temperature profile derived by the HASI instrument on board the Huygens probe. B: Pressure is calculated from the temperature profile by leveraging the law of hydrostatic equilibrium and approximating Titan's atmosphere as 100\% nitrogen. C: The integrated line-of-sight refractivity is found by stepping across 8000 km chords with step size = 1 km. At each step, refractivity $\nu$ is calculated, and the integrated line-of-sight refractivity is the sum of $\nu \cdot$ ds across the chord. D: Bending angles are calculated from the gradient of the phase delay screen. Orange vertical line in Panels B-D denotes Titan's surface. \label{fig:recipe}}
\end{figure}

Here is the step-by-step recipe for calculating the bending angles of rays that pass through Titan's atmosphere:

\begin{enumerate}

\item 	Begin with a temperature profile as a function of altitude. In this paper we use the Huygens/HASI profile as our starting reference for Titan’s atmosphere (Figure \ref{fig:recipe}A) \citep{fulchignoni_situ_2005}. While the HASI temperature profile lacks temporal and latitudinal resolution, it offers unparalleled altitudinal resolution and also has the advantage of being the only \textit{in situ} data of its kind collected in Titan's atmosphere. We also considered scaling the HASI profile by the latitudinal temperature differences simulated in the TAM GCM, but found the differences in the results to be negligible when compared to using the HASI profile without GCM scaling. We believe the HASI profile represents the best starting profile for Titan's atmosphere until additional highly altitudinally-resolved data can be collected.

\item	Calculate the pressure profile from $T(r)$ using the law of hydrostatic equilibrium and the assumption of a nitrogen atmosphere (Figure \ref{fig:recipe}B). There are differential and integral forms of the law of hydrostatic equilibrium; we use an integral form:

\begin{equation}
\ln(P) = - \int\frac{1}{H}dr
\label{hse}
\end{equation}

where $r$ is the radius, $P$ is the pressure, and $H$ is the scale height: $H(r) = \frac{kT(r)}{mg(r)} $. $k$ is Boltzmann's constant and $m$ is the molecular weight of nitrogen. A boundary condition is also required, such as $P$ at a certain $r$. Omitting methane leads to an overestimation of roughly 0.5\% in bending angles, assuming a constant 1.5\% concentration of methane at relevant occultation altitudes (extrapolated from the linearity of methane observed by HASI between 75 and 140 km).

\item	Calculate the number density profile, $N(r)$, from $T(r)$ and $P(r)$, using the ideal gas law.

\item	Account for zonal winds by distorting the number density field. The conversion of zonal winds to radial distortions is outlined in Section \ref{zw_to_dist}.

\item	Use the distorted number density field to calculate $\nu$, the refractivities. These are mainly proportional to the number density, with a very small dependence on wavelength. We calculate the refractivity through $N_2$ gas at standard temperature and pressure, then scale that refractivity by the ratio of the actual number density to Loschmidt's number, the number density at STP (Standard Temperature and Pressure: 273.15 K and 1 atmosphere).

\begin{equation}
\nu_{STP} = 29.06 \times 10^{-5} (1 + \frac{7.7 \times 10^{-3}}{\lambda^2}) 
\label{nu_stp}
\end{equation}

\begin{equation}
\nu(T,P) = \nu_{STP} \left( \frac{N(T,P)}{N_{LOS}} \right) 
\label{nu_tp}
\end{equation}

where $\lambda$ is the wavelength in $\mu$m and Loschmidt's number ($N_{LOS}$) is $2.687 \times 10^{19}$ particles per cm$^3$.

\item	Calculate the integrated line-of-sight refractivity for a given chord that traverses through Titan’s atmosphere (Figure \ref{fig:recipe}C). The most general way to do this is to construct chords that are perpendicular to the sky plane, break up the chords into small steps of length ds, evaluate the refractivity, $\nu$, at each step, and add up the $\nu$·ds products. We use step sizes of 1 km and a chord length of 8000 km, since number densities at distances of 4000 km (and beyond) from Titan’s center are negligible. While we considered the effect of Titan’s seasonally variable haze, we determined the effect was negligible due to the high altitudes that are probed by occultations and the wavelength range that the occultation was observed in. 

\item	Approximate Titan and its atmosphere as 2-D screens perpendicular to the observer’s line of sight, where each point on the screen has a light travel time that we calculate by summing $\nu \cdot ds$. The light travel time is proportional to the integrated line-of-sight refractivity. While Titan itself is decidedly 3-D, the 2-D screen approximation is perfectly acceptable – the bending angles we detect are only about 2 $\mu$rad, so that each detectable ray essentially travels straight through Titan’s atmosphere.

\item	Calculate bending angles from the gradient of the phase delay screen (Figure \ref{fig:recipe}D). In general, a wave propagating through a medium experiences bending angles equal to the gradient of the light travel times along axes perpendicular to the direction of travel. The 2-D sky-plane screen of bending angles lets us simulate resolved images of Titan. Specifically, we trace rays from the star through Titan’s atmosphere to the observer. This lets us construct the illumination around Titan’s limb at each instant in time. There are typically spots hovering over the limb at the locations of ingress and egress, but when the star is relatively near Titan’s center in the observer plane, the light distribution appears to move around the limb. We convolve the traced light locations with an appropriate PSF to produce an image that we can compare to the AO observations.

\end{enumerate}

\subsection{Variation in radial distortion profiles}

The earliest Titan winds occultation study \citep{hubbard_occultation_1993} constrained the zonal wind profiles by requiring them to be symmetric across the equator, ultimately deriving a zonal wind profile with peak speeds at latitudes of 65$^{\circ}$ N/S, lesser peaks at 30$^{\circ}$ N/S, and the lowest speed observed at the equator. Due to the Earth-based land geography of the 1989 occultation shadowpath, the observations from this occultation were concentrated in chords across Titan's southern hemisphere, but the equatorial symmetry constraint on zonal wind speeds meant that a zonal wind profile was proposed for all latitudes, even those not observed. 

About a dozen years later, \cite{bouchez_seasonal_2004} captured a double stellar occultation by Titan, obtaining the first spatially-resolved dataset of a Titan occultation. \cite{bouchez_seasonal_2004} argued that the equatorially symmetric constraint for Titan zonal wind profiles was unnecessary and that, based on Titan's atmospheric behavior, there is little reason to expect that the wind speeds would be symmetric about the equator throughout all of a Titan year. Subsequent observations and GCM simulations agree with the latter point -- most of the time, Titan hosts higher zonal wind speeds in the hemisphere experiencing autumn/winter. We do not constrain the symmetry of the zonal wind profiles and therefore consider Titan model atmospheres with substantially varying wind (and accordingly, distortion) patterns in the northern versus the southern hemispheres. Our forward model does not impose equatorial symmetry; however, we do assume that any given latitude will experience uniform winds throughout (i.e. we assume no longitudinal variations in wind speeds).

\subsection{Deriving level surface distortions from zonal winds}
\label{zw_to_dist}

\cite{hubbard_occultation_1993} call $\Delta r$ the ``deviation from spherical structure'', and their Fig. 34 shows the magnitude of $\Delta r$ as a function of latitude, ranging from 0 km at the poles to about 35 km at the equator. They express $r$ in terms of Legendre polynomials (their equations 20 and 23) and calculate $\Delta r$ by subtracting Titan’s polar radius. \cite{bouchez_seasonal_2004} derives a similar approximation for the ``shape of surfaces of constant pressure'', or $r(P,\phi)$, but uses an integral of zonal winds (integrated over latitude) instead of Legendre polynomials (see Eq. \ref{r_p_phi}, in which $P$ is pressure and $\phi$ is co-latitude). A key difference between the two expressions for level surfaces is that \cite{hubbard_occultation_1993} require that $\Delta r$ be symmetric about the equator, but \cite{bouchez_seasonal_2004} does not. Both require that the $\Delta r$ contours be perpendicular to the radius at the equator.

\cite{bouchez_seasonal_2004} refers to the surfaces of constant pressure as an ``altitude correction to the pressure structure of Titan’s atmosphere, due to the presence of zonal winds''. It may seem strange that these corrections only amount to a few tens of km, yet they can produce significant changes in the locations of the near- and far-limb spots during an occultation, as well as major changes in the shape of the central flash caustic in the middle of the shadow that Titan casts in the observer's plane. There is a reason that small distortions are so observable during occultations: it is because incoming rays are bent in a direction that is perpendicular to contours of constant density. Even small changes in $\Delta r$ can produce big changes in the orientation of isopycnal contours. This is especially apparent in central flash features, as depicted in the comparison for a simple oblate planet and a case based on a Titan-like distorted atmosphere (Fig. \ref{fig:sicardy_caustic}) \citep{sicardy_study_2022}.

In order to incorporate a zonal wind profile into the description of Titan’s atmospheric density structure, we follow the methodology conducted by \cite{bouchez_seasonal_2004} (Equations 4.16-4.26). We first apply radial distortions to a model Titan atmosphere (in this case, one derived from the HASI profile). To account for zonal winds, we begin by expressing Titan’s atmospheric rotation as the sum of the satellite’s rotation and the rotation due to zonal winds:

\begin{equation}
\omega(r,\phi) = \omega_s + \frac{V_w(r,\phi)}{r \sin \phi},
\label{rotation}
\end{equation}

\noindent where $V_w$ refers to zonal winds and $\omega_s$ refers to the satellite's angular rotation.

In hydrostatic equilibrium, the sum of the gravitational and centrifugal accelerations times the mass density of the gas is balanced by the pressure gradient. As gravitational acceleration vastly exceeds the radial term of centrifugal acceleration for zonal winds, we can disregard the centrifugal acceleration component when considering the partial derivative of pressure with respect to radius and express the relationship between radius, pressure, and colatitude as:

\begin{equation}
-g(\frac{\partial r}{\partial P})_\phi (\frac{\partial P}{r \partial \phi})_P = \omega^{2} r \sin \phi \cos \phi.
\label{partial_pressure_deriv}
\end{equation}

Rearranging Equation \ref{partial_pressure_deriv}, we get the partial differential equation describing shapes of constant pressure, which can then be integrated to calculate radius as a function of co-latitude. Radius $r$ and gravitational acceleration $g$ can be removed from the integral when the surfaces are nearly spherical, leading to Equation \ref{r_p_phi} for radius as a function of co-latitude in the presence of distorting zonal winds. A reference altitude $r(P,\phi_0)$ at reference co-latitude $\phi_0$ is used, in our case corresponding to the peak altitude from the ALMA contribution function. This equation is then applied in Step \#4 above. 
\begin{equation}
r(P, \phi) \approx r\left(P, \phi_0\right)\left(1-\frac{r\left(P, \phi_0\right)^3}{G M} \int_{\phi_0}^\phi \omega\left(P, \phi\right)^2 \sin \phi \cos \phi d \phi\right) .
\label{r_p_phi}
\end{equation}

\subsection{Quantifying refracted lightspot patterns}
\label{quant}

This paper compares the \emph{observed} light distribution around Titan's limb to \emph{simulated} distributions that would be observed if Titan's atmosphere had a particular 3D density structure. This paper is exclusively a forward modeling effort: we generate images of Titan during the occultation from GCM simulation snapshots of Titan or from atmospheres corresponding to the zonal winds observed by ALMA on 2022-09-06, then compare the synthetic limb light distributions to the observed distributions from Keck.

In general, concentrations of light on the limb appear where the projected line from the occulted star to the limb is perpendicular to contours of constant density. At the beginning of the occultation when the star has just been obscured by Titan, the lightspot appears along the limb at the location of ingress. The inverse is observed just prior to egress: the refracted lightspot is present at the point of egress (where the occultation chord and the edge of Titan's disk intersect). In the frames where the star was relatively close to the center of Titan's disk, refracted light patterns changed rapidly in an occultation-specific phenomenon known as a ``central flash''. Characterization of the comparatively faint refracted light spots required extensive processing to isolate them from Earth's atmospheric effects and the dominant light from Titan's disk. The deconvolution, stacking, and disk subtraction are described in detail in Section \ref{sec:process}. Examples of resulting isolated refracted lightspots are visible in Figure \ref{animated_occ}.

\begin{figure}[hbt!]
\centering
\includegraphics[width=18cm]{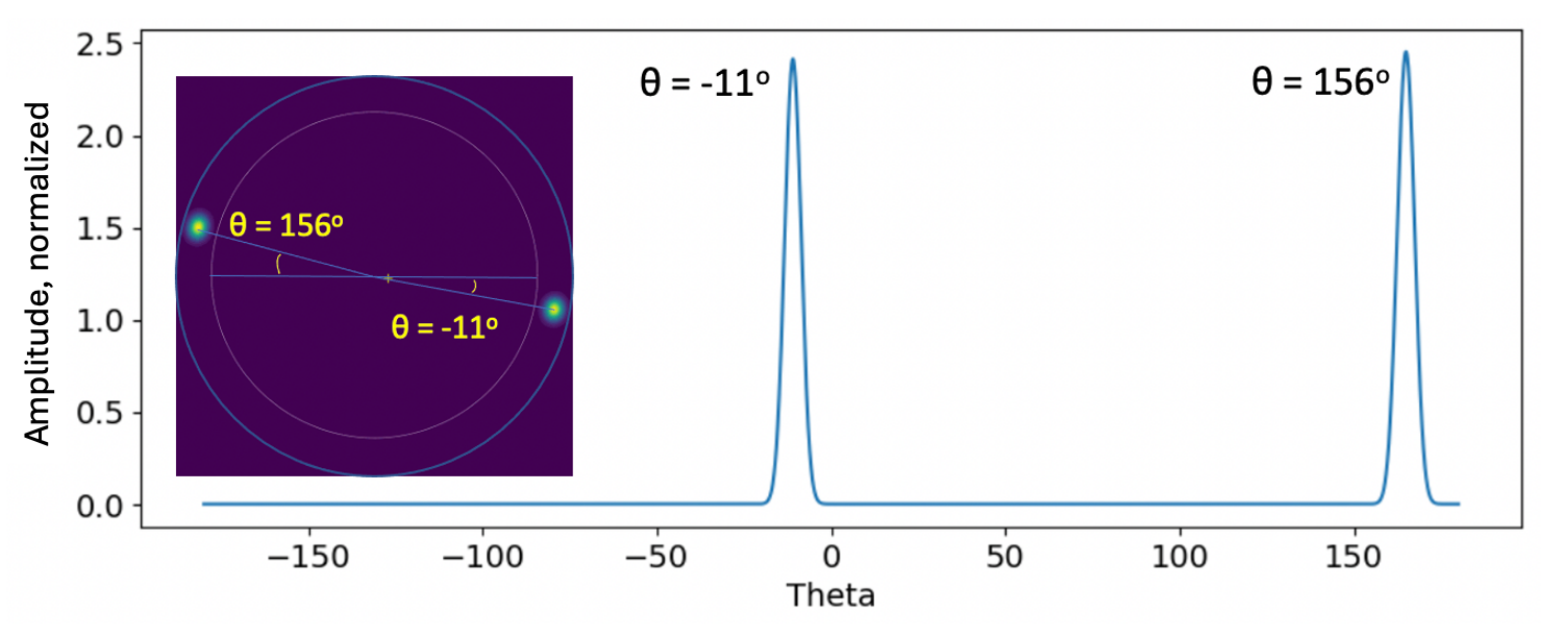}
\caption{Quantitative comparison of data and model can be achieved by treating Titan's disk as a circle and expressing the sum of the distribution of light around its limb as a function of angle $\theta$. This figure demonstrates this for a model frame near the center of the occultation. Both near and far limb spots are visible (inset), and when the sum of the light distribution around the limb is plotted, the peaks of the lightspots are found to be at $\theta = 156^{\circ}$ and $ \theta = -11^{\circ}$.\label{fig:summed_dist}}
\end{figure}

The final product of a simulated occultation using the forward model described in Section \ref{forward_model} and a model atmosphere can be compared to our data frame-by-frame. Quantitative comparisons were carried out by plotting the summed distribution of light around the limb in both data and model images as a function of $\theta$, with $\theta = 0^{\circ}$ in the positive x-direction (Figure \ref{fig:summed_dist}).

The duration of the occultation was just over six minutes - at the NIRC2 cadence of one frame per nine seconds, we collected 46 frames covering the entirety of the occultation. Due to the faintness of the star and variable interference from Earth's atmosphere, not all occultation data frames were usable. We identified an ``ingress lightspot" (appearing along the limb near the point of star ingress) in 22 of the 46 frames -- mostly in the first half of the occultation -- and an ``egress lightspot" (appearing along the limb near the point of star egress) in 18 of the 46 frames -- mostly in the second half of the occultation. Ten of the 46 frames had both near and far limb spots visible; these spanned the middle of the occultation. The ability of the model to match the data was quantified by finding the sum of squared errors (SSE) between data and model $\theta$ values for the ingress spot and the egress spot in data frames where the lightspots could be isolated. $\theta$ values for data and model were determined with Gaussian fitting \citep{matthew_newville_built-fitting_2024}.

\subsection{Forward model input from GCM zonal wind profiles}

We use a range of GCM simulation snapshots and the near-concurrent ALMA zonal wind data for our forward model atmosphere profile sample space. As outlined in Section \ref{forward_model}, a 2D temperature profile can be converted to a 2D density profile, which allows calculation of refractivities and ultimately bending angles for refracted starlight. 

\begin{figure}[hbt!]
\centering
\includegraphics[width=18cm]{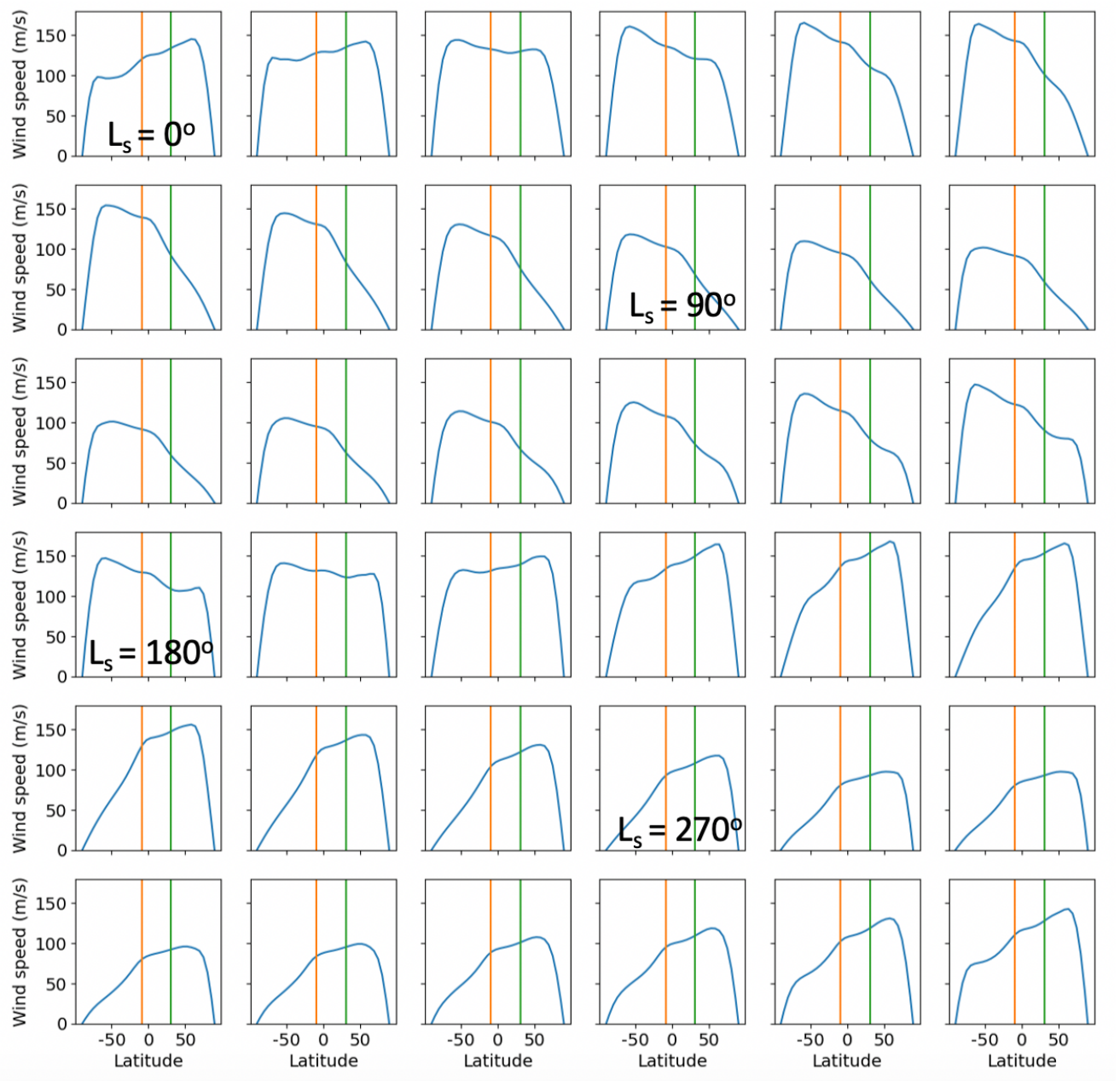}
\caption{Snapshots of simulated zonal wind profiles from TAM at an altitude $z = 307$ km. $L_s$ values of 0$^{\circ}$, 90$^{\circ}$, 180$^{\circ}$, and 270$^{\circ}$ correspond to northern vernal equinox, summer solstice, autumnal equinox, and winster solstice, respectively. Wind trends are most dynamic around the two equinoxes. Orange line represents the latitude of ingress ($-9^{\circ})$ and green line represents latitude of egress ($31^{\circ})$.\label{fig:zw_all}}
\end{figure}

We used 36 snapshots of 2D zonal wind profiles over a Titan year derived from simulations with TAM. The limited vertical resolution in TAM 2D density profiles means the resulting refractivity profiles lack the necessary resolution across occultation-relevant altitudes. Instead of interpolating across the GCM 2D density profiles and using these profiles to construct our refractivity screens and bending angles, we chose to apply the radial distortions derived from zonal wind profiles in each GCM snapshot (distortion calculations outlined in Section \ref{zw_to_dist}) to a refractivity profile calculated from the HASI temperature profile. Radial distortions across the range of altitudes in the TAM GCM were calculated as per Eq. \ref{r_p_phi} and the refractivity profile from HASI was stretched accordingly. We use the integral form of hydrostatic equilibrium in our forward model exploring modifications to surfaces of level pressure exerted by zonal winds. The wind profiles extracted from the TAM simulations are depicted in Figure \ref{fig:zw_all} for a reference altitude (307 km). While 36 snapshots are displayed, we note that the shape of the zonal wind distribution does not change in every snapshot. Changes are most readily observed around the equinoxes ($L_s = 350^{\circ}$ to $L_s = 40^{\circ}$ and $L_s = 170^{\circ}$ to $L_s = 220^{\circ}$); between equinoxes, the change between snapshots is primarily in the magnitude of the zonal wind speeds. The observed occultation occurred around $L_s = 160^{\circ}$, a few Earth years (0.07 Titan years) before Titan's northern autumnal equinox in 2025.

\begin{figure}[hbt!]
\centering
\includegraphics[width=18cm]{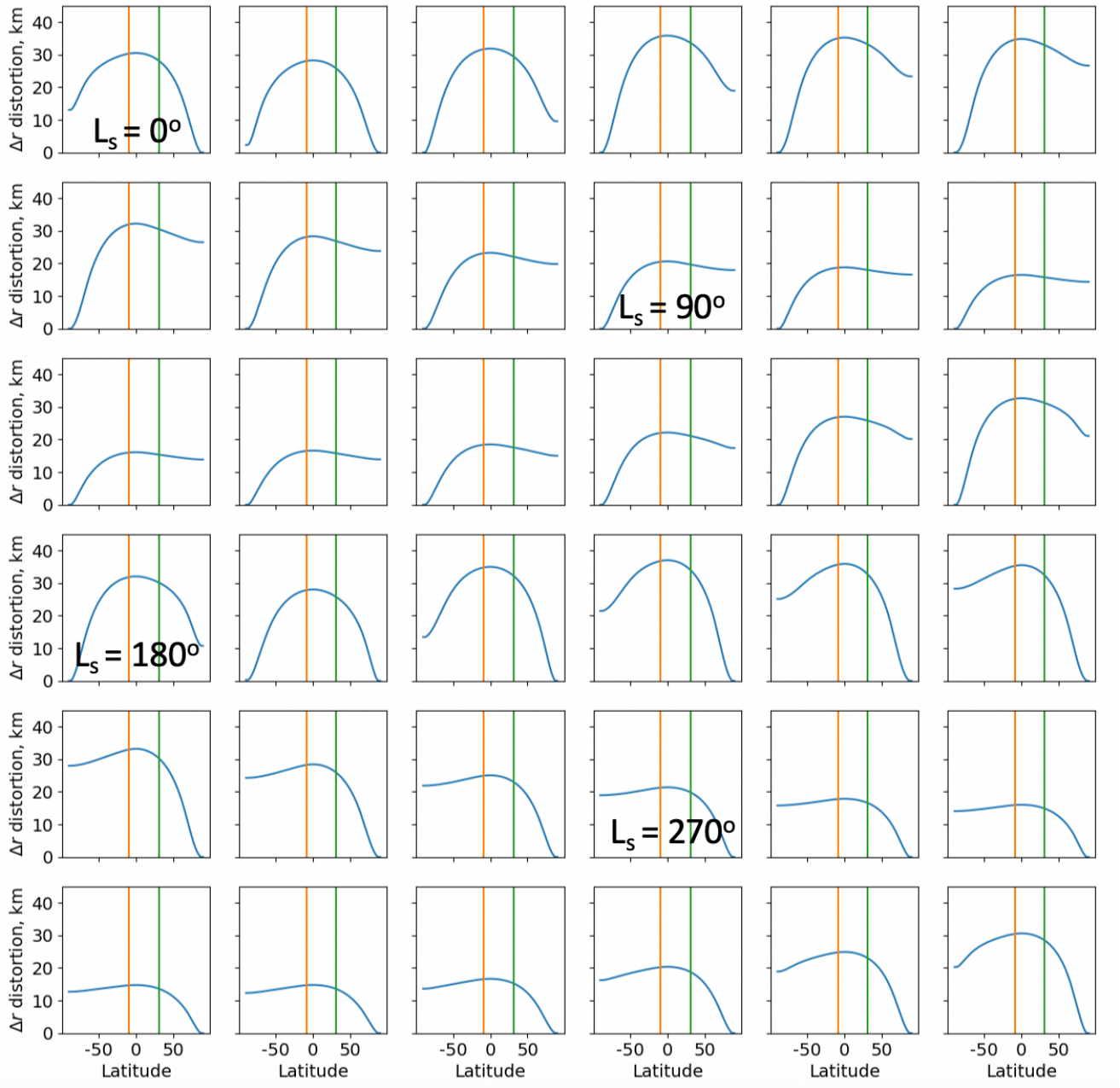}
\caption{The radial distortions (contours of level pressure) corresponding to the 36 zonal wind snapshots in Figure \ref{fig:zw_all}. Orange line represents the latitude of ingress ($-9^{\circ})$ and green line represents latitude of egress ($31^{\circ})$.\label{fig:deltar_all}}
\end{figure}

The radial distortions corresponding to the GCM snapshot zonal wind speeds are depicted in Figure \ref{fig:deltar_all}. While zonal wind profiles underlie the observed radial distortions, it is the radial distortions that ultimately affect the pattern of the refractivity screens and, by extension, bending angles. Latitudinally symmetric GCM profile snapshots (e.g., at $L_s = 10^{\circ}$ and $L_s = 190^{\circ}$) lead to latitudinally symmetric distortion profiles, although the majority of profiles are asymmetric about the equator. In addition to the 36 GCM zonal wind profiles here, we ran the forward model with both doubled and halved zonal wind speeds from each of the GCM snapshots to determine the effect of increased and decreased atmospheric distortion on the correspondence between the data and the model.

\section{Results}

\subsection{Retrieved zonal wind field using ALMA}

\begin{figure}[h!]
\centering
\includegraphics[width=7cm]{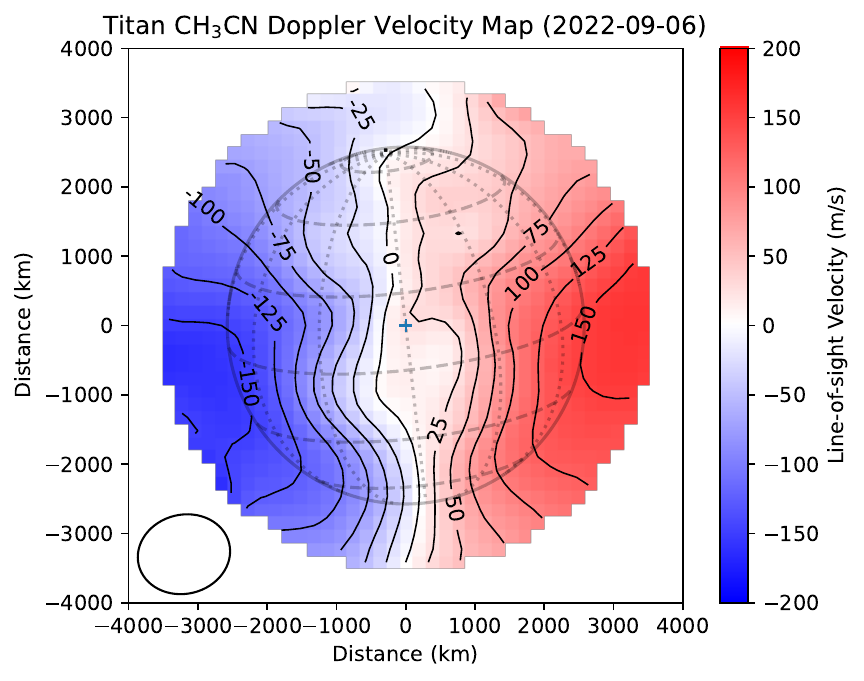}
\includegraphics[width=7cm]{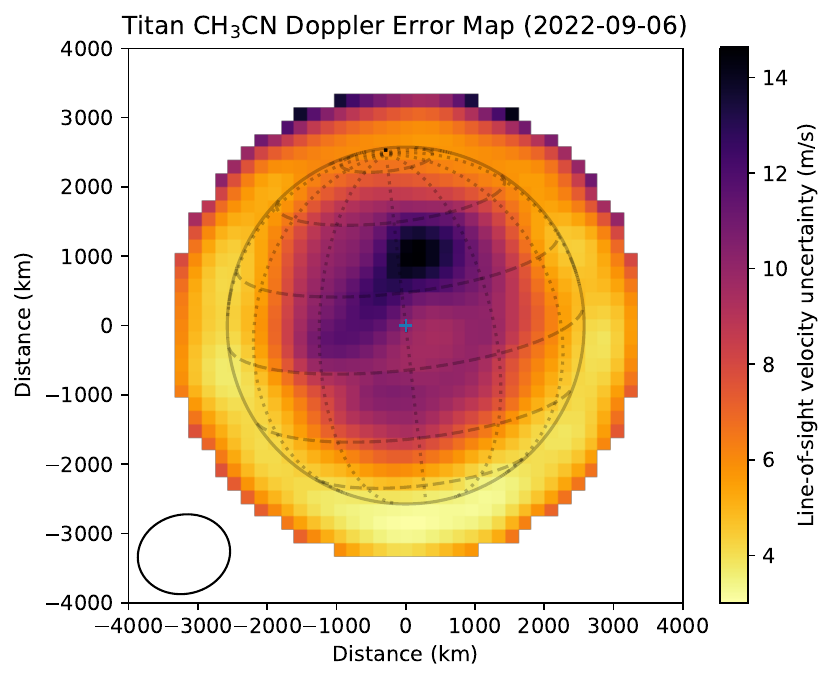}
\caption{Left: The color map shows line-of-sight CH$_3$CN Doppler shifts, averaged across all observed transitions, overlaid on a Titan wireframe grid at the epoch of our ALMA observations. Velocity contours (labeled in units of m\,s$^{-1}$) are also overlaid. Right: Error map showing $1\sigma$ statistical uncertainties on the Doppler shift for each pixel. The spatial resolution (elliptical beam FWHM) is indicated in the lower left.\label{fig:doppler}}
\end{figure}

\begin{figure}[h!]
\centering
\includegraphics[width=9cm]{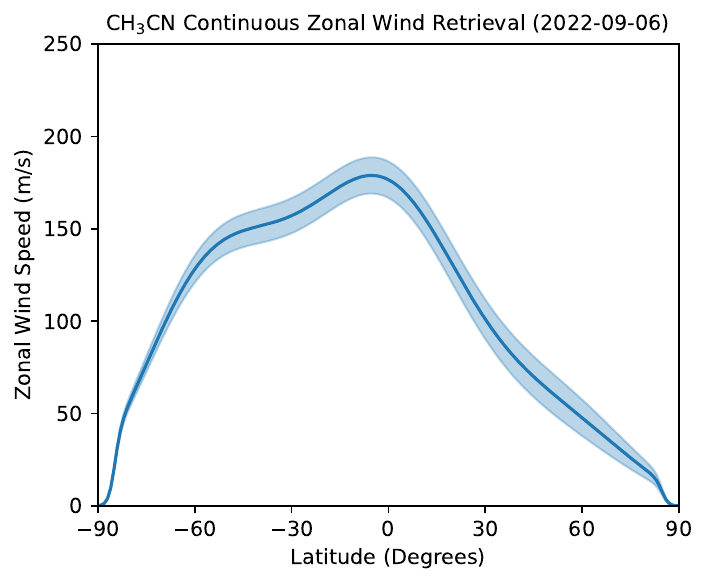}
\caption{Zonal wind velocity profile ($v(l)$) and $1\sigma$ error envelope retrieved from our ALMA CH$_3$CN observations on 2022-09-06, using a continuously-variable parameterization with latitudinal sampling of 30$^{\circ}$. \label{fig:almawindlat}}
\end{figure}

The final Doppler map (averaged over all eight CH$_3$CN lines) from our ALMA observations is shown in Figure \ref{fig:doppler}, along with the $1\sigma$ uncertainty map for each pixel. The data have been masked within a circular aperture of radius 3600~km for display purposes. The resulting best-fitting latitudinal wind profile is shown in Figure \ref{fig:almawindlat}. This wind profile corresponds to the CH$_3$CN contribution function depicted in Figure \ref{fig:contrib}.

\subsection{Occultation forward model with GCM and ALMA input atmospheres: comparison to data}

\subsubsection{Data versus model: overall}

\begin{figure}[hbt!]
\centering
\includegraphics[width=17cm]{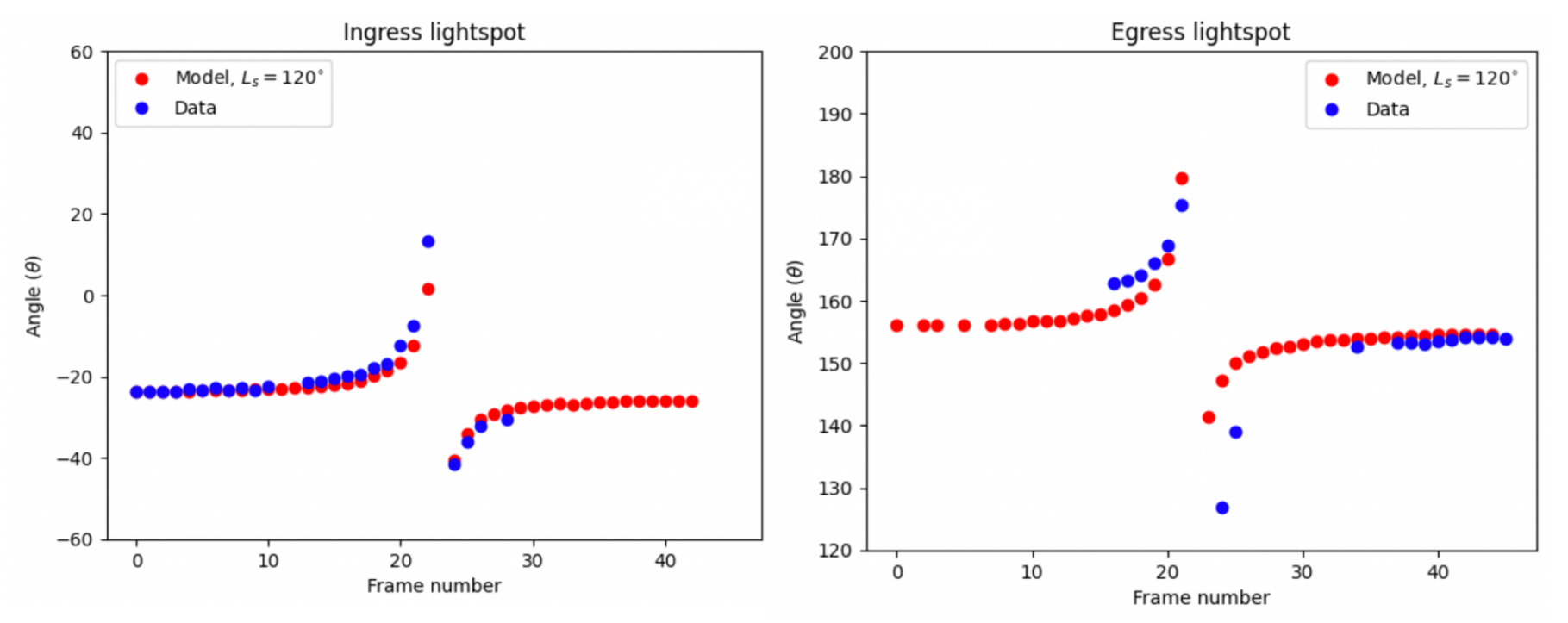}
\caption{Best-match distributions of the angles of the refracted lightspots visible during occultation on the ingress and egress sides of Titan's disk. Data are depicted in blue and the model simulations with a starting atmosphere from the TAM simulation at $L_s = 120^{\circ}$ are depicted in red. \label{fig:combine_fit}}
\end{figure}

Of the input zonal wind profiles surveyed (36 GCM snapshots and 1 ALMA profile), the overall best match to the occultation data was for the input GCM atmosphere corresponding to $L_s = 120^{\circ}$, closely followed by those corresponding to $L_s = 110^{\circ}$, $L_s = 130^{\circ}$, and $L_s = 100^{\circ}$. $L_s =120^{\circ}$ is $40^{\circ}$ - or roughly half of a Titan season - from the time of the occultation ($L_s 160^{\circ}$), and both profiles are from Titan's northern summer. The profiles for $L_s = 120^{\circ}$ and $L_s = 160^{\circ}$ are similar in shape and vary only in peak zonal wind magnitude. The degree to which the occultation simulations generated by various input zonal wind profiles matched the data was measured by calculating SSE values for data versus model as described in Section \ref{quant} (Table \ref{tab:chisq}; full table of SSE values can be found in the Appendix (Table \ref{tab:si_table1})). We note that the zonal wind profiles from these four GCM snapshot simulations are nearly identical (Figure \ref{fig:zw_all}) and that variation between the matches for these profiles was low. The data-model comparison is shown in Figure \ref{fig:combine_fit}.

\begin{table}[ht]
\begin{center}
\caption{SSE table of closest model-data matches for original amplitude starting profiles.\label{tab:chisq}}
\begin{tabular}{||c | c | c | c | c | c||} 
 \hline
 Combined & SSE & Ingress & SSE & Egress & SSE\\ 
 \hline\hline
 $L_s 120^o$ & 570.5 & ALMA & 42.6 & $L_s 120^o$ & 496.7 \\ 
 \hline
  $L_s 110^o$ &  573.2 & $L_s 10^o$ & 52.8 & $L_s 110^o$ & 499.9 \\
 \hline
  $L_s 130^o$ & 576.7 & $L_s 190^o$ & 53.8 & $L_s 130^o$ & 504.3 \\
 \hline
  $L_s 100^o$ & 584.1 & $ L_s 70^o$ & 56.5 &  $L_s 100^o$ & 512.6 \\
  \hline
\end{tabular} \\
\flushleft\footnotesize{Sum of squared errors table of the closest model-data matches for both ingress and egress spots combined, ingress spots alone, and egress spots alone. Only the original 37 input profiles (36 GCM snapshots and 1 ALMA profile) are included. See Appendix, Table \ref{tab:si_table1} for all SSE values for original magnitude profiles.}
\end{center}

\end{table}

\begin{figure}[hbt!]
\centering
\includegraphics[width=17cm]{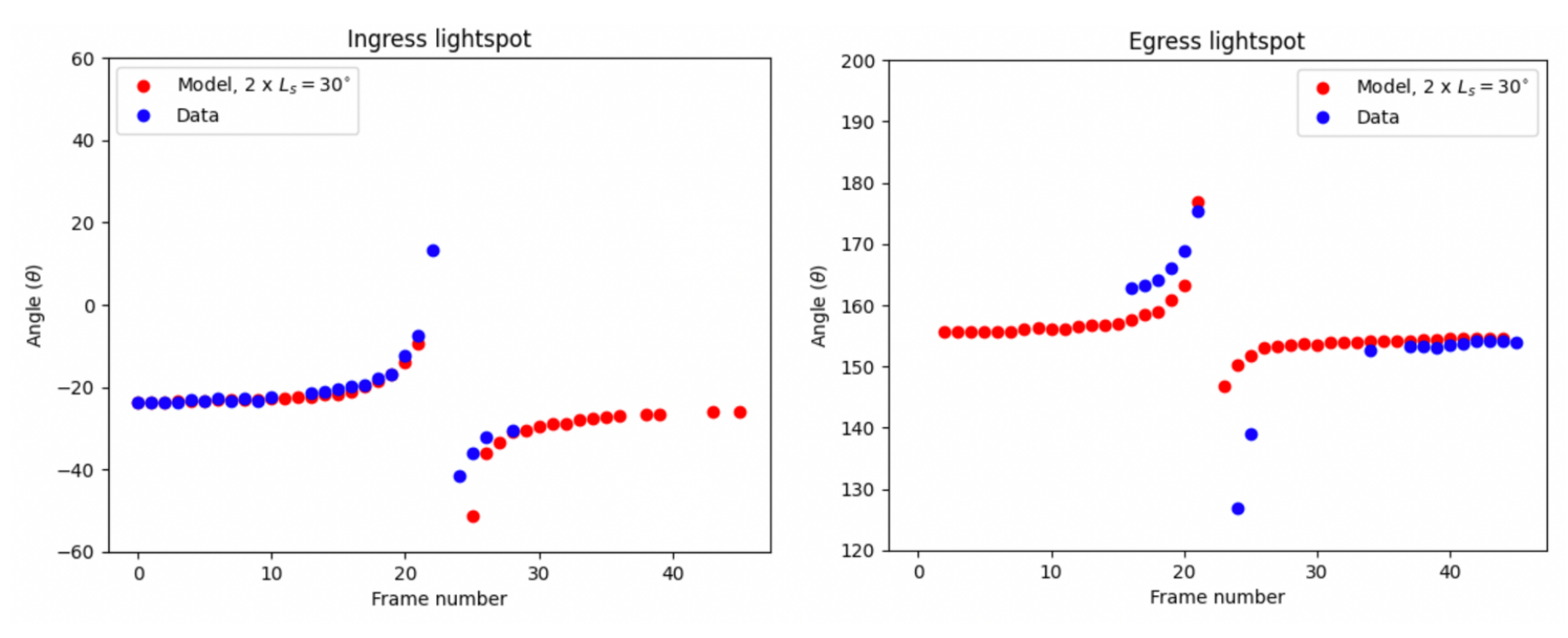}
\caption{Distributions of the angles of the refracted lightspots visible during occultation on the ingress and egress side of Titan's disk. Data are depicted in blue and the model simulations (with an atmosphere corresponding to double the GCM wind speeds) at $L_s = 30^{\circ}$ are depicted in red. \label{fig:combine_double_fit}}
\end{figure}

We considered how increasing or decreasing the GCM wind speed profiles by a factor of two (and thus accordingly increasing or decreasing the distortions of surfaces of level pressure, as described in Section \ref{zw_to_dist}) would affect the matches. When evaluated across all residuals (ingress and egress lightspots, all frames), it was apparent that doubled wind speeds led to poorer matches across all simulations. An example poor match to a doubled GCM profile is depicted in Figure \ref{fig:combine_double_fit}. We note that worse matches are not necessarily universally worse: rather, the comparison between model and data for the ingress lightspot in the first half of the occultation (Figure \ref{fig:combine_double_fit}, frames 1-23 of left panel) represent a better match than the best overall match depicted in Figure \ref{fig:combine_fit}. However, the trends for the egress lightspot as well as the latter frames for the egress lightspot denote a marked departure from the trends observed in the data, and the overall matches of the doubled zonal wind profiles from the GCM simulations perform more poorly than those of the non-doubled GCM-simulated zonal wind profiles. Best matches for the expanded set of forward model input zonal wind profiles are presented in Table \ref{tab:ext_chisq} (all SSE values listed in Appendix Tables \ref{tab:si_table2} and \ref{tab:si_table3}).

\begin{table}[ht]
\begin{center}
\caption{SSE table of closest model-data matches for all starting profiles.\label{tab:ext_chisq}}
\begin{tabular}{||c | c | c | c | c | c||} 
 \hline
 Combined & SSE & Ingress & SSE & Egress & SSE\\ 
 \hline\hline
 0.5 $\times L_s 110^o$ & 512.8 & 2 $\times L_s 270^o$ & 26.0 & 0.5 $\times L_s 110^o$ & 428.1 \\ 
 \hline
  0.5 $\times L_s 130^o$ &  514.4 & 2 $\times L_s 240^o$ & 29.0 & 0.5 $\times L_s 130^o$ & 430.5 \\
 \hline
  0.5 $\times L_s 80^o$ & 516.4 & 2 $\times L_s 340^o$ & 30.3 & 0.5 $\times L_s 120^o$ & 433.6 \\
 \hline
  0.5 $\times L_s 120^o$ & 518.0 & 2 $\times L_s 250^o$ & 30.8 & 0.5 $\times L_s 80^o$ & 436.3 \\
  \hline
\end{tabular}

\flushleft\footnotesize{Sum of squared errors table of the closest model-data matches for both ingress and egress spots combined, ingress spots alone, and egress spots alone. Input profiles include the original 36 GCM and 1 ALMA inputs (that are shown above), and the same 37 profiles with both halved and doubled magnitudes (total of 111 input profiles). See Appendix, Tables \ref{tab:si_table2} and \ref{tab:si_table3} for all SSE values for halved and doubled magnitude profiles.}
\end{center}
\end{table}

\begin{figure}[hbt!]
\centering
\includegraphics[width=17cm]{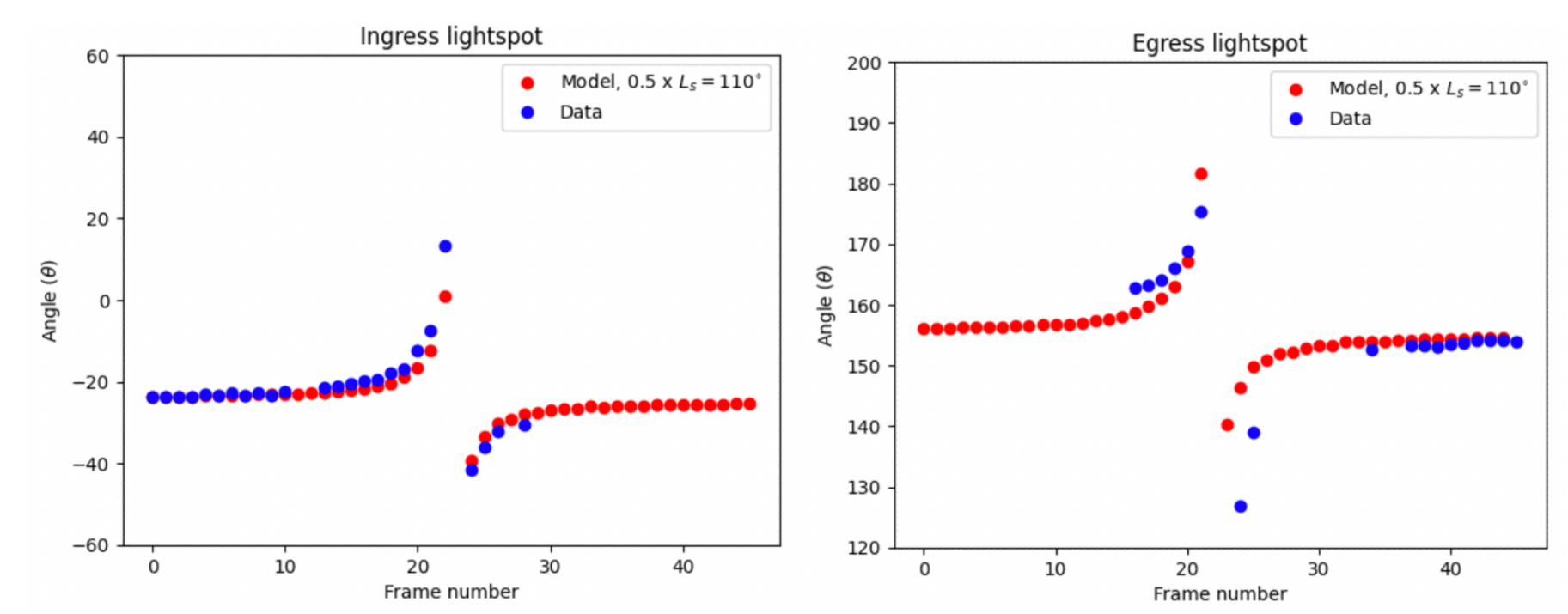}
\caption{Distributions of the angles of the refracted lightspots visible during occultation on the ingress and egress side of Titan's disk. Data are depicted in blue and the model simulations (with an atmosphere corresponding to half the GCM wind speeds) at $L_s = 120^{\circ}$ are depicted in red. \label{fig:combine_half_fit}}
\end{figure}

Halving the wind speed magnitudes from the GCM led to slightly improved matches, although the differences in matches compared to the original magnitudes were far smaller than those observed between doubled and original magnitudes. Halving the wind speed profiles did not improve the matches for the ingress spot, but generally led to better matches for the egress spot in many cases (e.g., Figure \ref{fig:combine_half_fit}). In the set of simulations with halved magnitudes, $L_s = 120^{\circ}$ was no longer the best match -- there were three $L_s$ values that produced closer matches to the data ($L_s = 110^{\circ},L_s = 130^{\circ}$, and $L_s = 80^{\circ}$) (Table \ref{tab:ext_chisq}). 

Due to Titan's north pole angle on the night of the observations ($\theta = 6.522^{\circ}$) and the trajectory of the occulted star, the latitudes of ingress and egress were significantly different from each other ($-9.2^{\circ}$ at ingress and $31.2^{\circ}$ at egress). We note that the match between data and model is significantly better for the ingress lightspot than it is for the egress lightspot, and that different atmospheric profiles emerge as the best match when each lightspot is considered individually. The latitude of ingress ($-9.2^{\circ}$) is closer to the equator than that of egress ($31.2^{\circ}$), and the asymmetry of the matches suggest that the GCM zonal wind profiles more closely match the data nearer to the equator. We note that the radial distortion profile is constrained by a slope of zero at the equator and monotonically decreases from equator to pole on both sides. This meant that for the forward models tested, the slopes for latitudes near the equator were nearer to zero and less variant than those farther from the equator. The monotonic and equatorial constraints on the radial distortion profiles come from the fact that atmospheric zonal flow will cause centrifugal acceleration directed outward radially, parallel to the equatorial plane. This force is accordingly balanced by an equator-pointing positive pressure gradient. Thus, surfaces of constant pressure from equator to both poles slope downward monotonically \citep{bouchez_seasonal_2004}. 

After determining the combined best match, we probed each individual lightspot to see what distinct trends appeared at ingress and egress.

\subsubsection{Data versus model: Ingress lightspot}

\begin{figure}[hbt!]
\centering
\includegraphics[width=17cm]{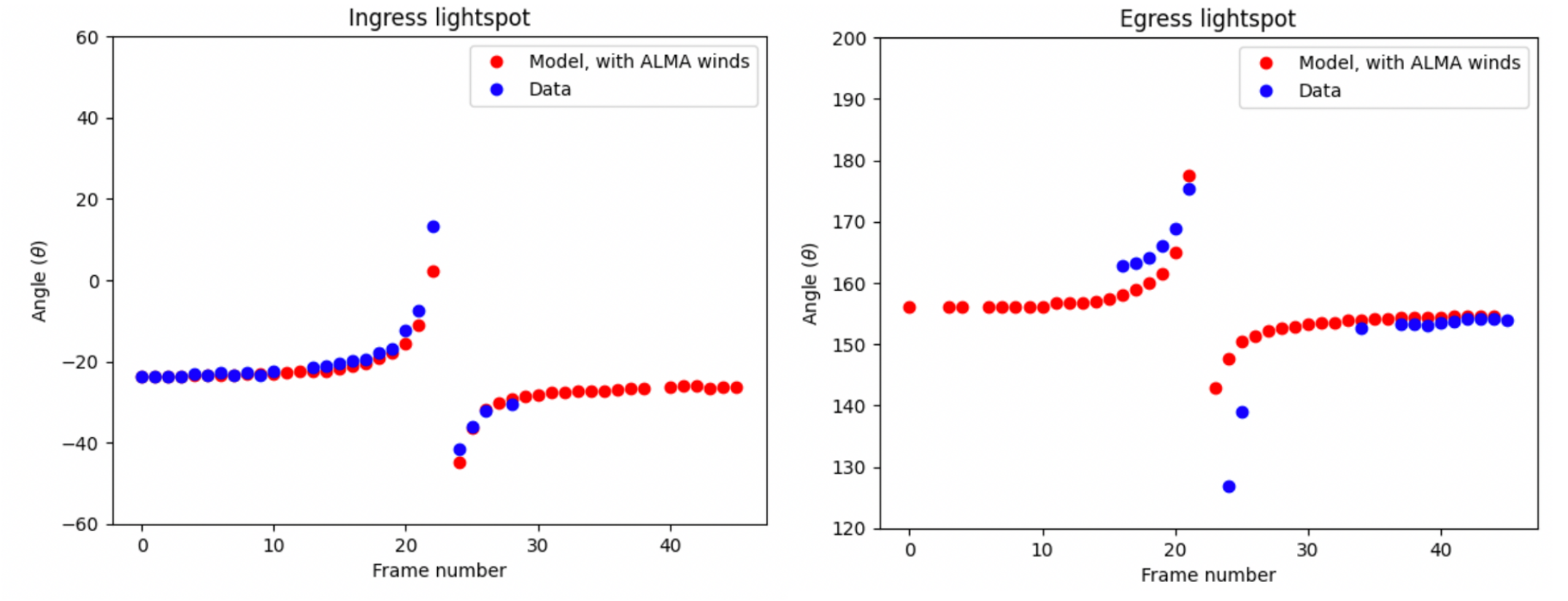}
\caption{Distributions of the angles of the refracted lightspots visible during occultation on the ingress and egress side of Titan's disk. Data are depicted in blue and the model simulations are depicted in red. Model simulations correspond to a starting zonal wind profile derived from the ALMA data. \label{fig:ingress_fit}}
\end{figure}

Of the original input zonal wind profiles surveyed (36 GCM snapshots and 1 ALMA profile), the model that best matches the data for the ingress lightspot was the ALMA input atmosphere, followed by $L_s = 10^{\circ}$ and $L_s = 190^{\circ}$, the two zonal wind profiles corresponding to the most symmetric radial distortions (Figure \ref{fig:deltar_all}). Data/model matches are shown in Figure \ref{fig:ingress_fit} and SSE values are in Table \ref{tab:chisq}. 

Although doubling the wind speeds did not lead to improved matches overall, several of the doubled zonal wind profiles from the GCM simulations did lead to a better match for the ingress lightspot. Doubling the zonal wind profile from the GCM simulation at $L_s = 270^{\circ}$ led to the closest match of all starting atmospheric profiles tried, exceeding even the ALMA starting profile in quality of the match. While the best match for the combined ingress and egress lightspot favored zonal wind profiles with lower peak speeds, this restriction was not present when the ingress spot was considered alone. 

\subsubsection{Data versus model: Egress lightspot}

The egress lightspot trends match those of the best match for the two lightspots combined. The best match for the egress lightspot is found at $L_s = 120^{\circ}$, with $L_s = 100-110^{\circ}$ and $L_s = 130-140^{\circ}$ close behind. The match is visualized on the right side of Figure \ref{fig:combine_fit}. All of these GCM simulated zonal wind profiles are similar in shape and magnitude, with their peak zonal winds being low compared to other snapshots in the course of a Titan year. Doubling the zonal wind profiles led to substantially poorer matches, suggesting that there is some dependence of the match of the egress lightspot on the peak zonal wind (favoring lower values).

\section{Discussion}
   
\subsection{On the better model match to the ingress than egress data}

We propose the GCM zonal wind simulations for $L_s = 120^{\circ}$ as the overall closest match among the GCM simulations for the observed occultation. The GCM zonal wind profile simulated for $L_s = 120^{\circ}$ has the same basic shape as $L_s = 110^{\circ}$ and $L_s = 130^{\circ}$. The primary difference is that $L_s = 120^{\circ}$ has the lowest peak zonal wind speeds within that time period of Titan's year, with the highest speeds reaching just under 100 m/s. When forward models were run with halved wind speeds, we determined that these half-magnitude simulations match the data slightly better overall compared to the original magnitudes. In both original magnitude and half-magnitude cases, the best-matching profiles had the same shape (peak winds at southern latitudes). The correlation between data-model match and peak wind speed magnitude was only observed in the case of the original magnitude simulations.

The best match was driven by the degree of data-model overlap observed for the egress lightspot (at latitude of $31^{\circ}$), which yielded higher residuals (i.e., poorer matches) than those observed for the ingress lightspot. It is additionally possible that a zonal wind profile that varies distinctly from all simulated GCM profiles would provide a better match to the data. 

It is noteworthy that the improved match with lower peak zonal wind speed observed for the egress lightspot is not observed for the ingress lightspot. The best match for the ingress lightspot was observed with the ALMA data with a peak wind speed just under 200 m/s, higher than nearly all of the zonal wind profiles simulated by the GCM. Furthermore, when the GCM zonal wind speeds were doubled and the resulting profiles were inputted into the forward model, we found that some of these profiles emerged as better matches even than the ALMA data, and the ones that were the best matches had high peak wind speeds and radial distortions.

\subsection{Level Surface Contours and the Mid-Event Frame: Evidence for a Southern Stratospheric Jet}

The 2022-09-05 occultation was special because the Maunakea observatories were situated very close to the central chord. Images taken when the star is behind the center of Titan's disk are very important due to the geometry of refracted rays around the entire limb. We discuss that geometry now in the context of zonal winds, the corresponding level surfaces in the atmosphere ($\Delta r$), and what we can learn about the $\Delta r$ contours from the distribution of light around the limb.

\begin{figure}[hbt!]
\centering
\includegraphics[width=10cm]{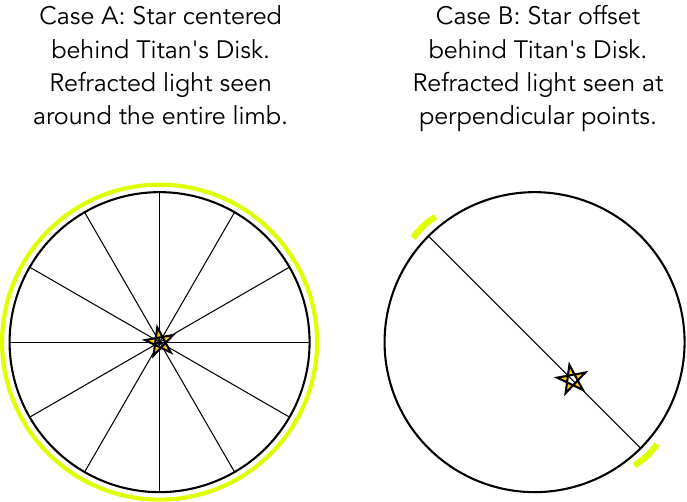}
\caption{Limb light distributions for a circular Titan. Case A (left panel) shows Titan as seen by an observer at mid-event with the star directly behind Titan's disk. If Titan's atmosphere is circularly symmetric, there will be a ring of light around the entire limb because the projected lines from the star to the limb are perpendicular to the isopycnal lines all around the disk. In Case B, the star is offset from the center, and there are only two regions where the projected lines from the star are perpendicular to the isopycnal lines. \label{fig:CircularCase_PerpendicularLimbRegions}}
\end{figure}

We start with the fact that the bending angle is proportional to the gradient of the line-of-sight integrated density field. This means that rays from a distant star will bend in a direction that is perpendicular to contours of constant density. If Titan were completely spherical, the occultation shadow in the observers' plane would have a sharp bright spot in the middle from all of the rays that pass through a certain annulus above Titan's limb (the annulus where the bending angle is equal to $Radius_{Titan}/Distance_{Earth-Titan}$, or $\sim 2 \mu{rad}$). Observers at the center of the shadow would see a ring of refracted starlight around Titan's entire circular disk, because from their point of view, the projected radial lines from the star to the isopycnal contours are perpendicular around Titan's entire limb (Fig. \ref{fig:CircularCase_PerpendicularLimbRegions}). That is the value of an AO-capable site near the central chord -- observers get to see where level surface contours are perpendicular to radial lines around Titan's entire disk. When the star is \emph{not} directly behind the center of Titan's disk (as is the case for most or all of an occultation), there will be two regions around the limb where the perpendicular conditions hold.

\begin{figure}[hbt!]
\centering
\includegraphics[width=10cm]{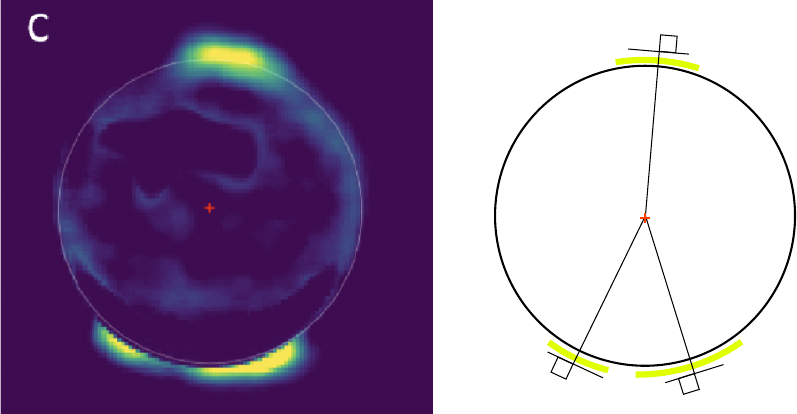}
\caption{The left panel shows the resolved image of Titan taken closest the mid-event, when the star is nearly directly behind Titan's disk. The right panel is a diagram to show regions where the $\Delta r$ level surface is perpendicular to the radial lines from star. \label{fig:NonCircularCase_PerpendicularLimbRegions}}
\end{figure}

What if the star is directly behind Titan's disk but Titan's projected atmosphere is \emph{not} circular? In this case the arcs of refracted light show exactly where the isopycnal lines are perpendicular to the radial lines from the star. A rough interpretation of Frame B in Figure \ref{animated_occ} is that there is a small perpendicular region near the north pole plus two broader regions around the south pole (Fig. \ref{fig:NonCircularCase_PerpendicularLimbRegions}). The implication is that Titan's atmosphere is shaped like an egg with the pointy end near the north pole, a pattern that is consistent with stronger winds in the southern hemisphere. The near- and far-limb spots move dramatically in the few frames on either side of mid-event: these frames also constrain regions where the projected line from the star must be perpendicular to isopycnal lines. A high-cadence series of images near mid-event can map out the slopes of the $\Delta r$ surface over most of Titan's limb. In theory, the slopes could be integrated to get the $\Delta r$ surface itself, from which a profile of zonal winds could then be recovered. The NIRC2 data rate was one frame every 9 s, which was limited in part by the speed of the network in saving frames to disk. Still, the distribution of light around the limb in the five or six mid-event frames will be useful in constraining zonal winds, especially near the poles. We note that neither the ALMA zonal wind profile nor any of the GCM profiles produce refracted light distributions like Frame B of Fig. \ref{animated_occ}.

\subsection{Comparison between GCM, occultation, and ALMA zonal winds}

Stellar occultations, ALMA CH$_3$CN Doppler shifts, and TAM GCM simulations probe overlapping altitudes in Titan's atmosphere: occultations are sensitive to roughly 180-500 km, CH$_3$CN Doppler shift data are most sensitive to 200-400 km, and TAM GCM simulations provide predictions of wind speeds up to 650 km altitude. Overall, the GCM snapshots with higher wind speeds at southern latitudes seem to provide the best match, indicating that the GCM predictions for the solar longitude corresponding to our observations are reasonable. We additionally find that lowering the peak wind speed of the GCM profiles led to slightly better data-model matches, although the profiles that best match the ingress spot were different from those that best match the egress spot. The forward model resulting from the ALMA wind profile matches the data for the ingress spot better than any of the GCM model simulations, although its match to the egress spot was poorer. The shape of the derived ALMA wind profile did not closely match any of the GCM wind profiles for the surrounding $L_s$ values, which had peak wind speeds at latitudes farther from the equator. We note that the implementations of ALMA and GCM profiles into our forward model varied slightly due to GCM simulations being altitudinally resolved while ALMA retrieved wind speeds are averaged over an altitude contribution function (Figure \ref{fig:contrib}). Thus, a forward model implementation with a GCM starting profile incorporates wind speeds that vary with altitude, which may account for some of the differences observed when employing the two different types of starting profiles.

We considered results from three distinctly different methods of studying zonal winds: computational simulations from General Circulation Models, characterization of radial distortion profiles and, accordingly, zonal wind profiles via the behavior of refracted starlight, and zonal winds derived from ALMA Doppler shifts of specific molecules in Titan’s atmosphere. The GCM affords the ability to define and refine input parameters, specify latitudes, and simulate profiles across an entire Titan year. However, as the simulations consist of free-running numerical simulations, they do not necessarily probe the real system like the other two techniques discussed here.

Zonal wind profiles from ALMA data of Doppler shifts of a specific molecule probe a large range of altitudes in Titan's atmosphere. Spatial resolution is driven by the configuration of the ALMA dishes on the night of the observation and CH$_3$CN (the molecule probed in the results presented here) has a contribution function that reaches higher altitudes than either the TAM simulations or stellar occultations, albeit at limited sensitivities above 400 km (Figure \ref{fig:contrib}). As the thermospheric equatorial jet proposed by \cite{lellouch2019intense} has been shown to extend down into the mesosphere \citep{cordiner_detection_2020}, it is possible that the jet has a small influence on the overall ALMA wind profile, which would be less prominent in the occultation data and GCM simulations due to their lower maximum altitudes. As the ALMA wind profile is not vertically resolved, isolating the contribution of the equatorial jet from the contributions of winds at lower altitudes that contribute more heavily to the CH$_3$CN profile is not possible. 

Stellar occultation observations with adaptive optics-equipped telescopes present a unique chance to directly image refracted starlight as it moves around Titan's limb. The combination of starting GCM-derived wind profiles and the HASI temperature profile allowed vertical resolution to be maximized across the sensitive range. However, the process of comparing refracted lightspots to potential atmospheres that could generate such patterns is cumbersome and computationally expensive, making inversion with ray-tracing time-intensive. Development of faster pipelines has not been a priority due to the rarity of observable occultations and datasets; the 2022 Titan occultation is only the second one to be observed with adaptive optics and the first one in more than 20 years. With this dataset, an additional recent Titan occultation dataset, and an upcoming 2026 Titan occultation, we are hopeful for increased attention toward this field in the coming years.

\section{Conclusions}

We present observations of a stellar occultation by Titan obtained with Keck on 2022-09-05, together with ALMA data of the satellite one day later. We use a forward model to simulate the occultation, and to compare the stellar occultation data to the ALMA profile and to 36 GCM profiles spanning an entire Titan year, and the same set of 37 profiles with both doubled and halved zonal wind speeds. We find that overall, the GCM profile corresponding to a solar longitude of $120^{\circ}$ (roughly $ 40^{\circ}$ before the time of the occultation) matches the data best, with similarly shaped profiles with neighboring $L_s$ values following close behind (Table \ref{tab:chisq}). The zonal wind profile (and corresponding radial distortion profile) for $L_s = 120^{\circ}$ is not substantially different than the neighboring $L_s$ values, and it has a low peak zonal wind and peak radial distortion value compared to many of the other $L_s$ values from the GCM simulations. The peak zonal wind is found in the southern hemisphere, which is consistent with GCM predictions given the Titan's season at the time of the occultation. The findings of this paper may aid future studies of Titan's climate, chemistry, and long-term atmospheric evolution, as well as constrain and inform findings of the upcoming \textit{Dragonfly} mission.

There was a notable disparity between the matches of the ingress lightspot and the egress lightspot to the data, with the best model match to the ingress lightspot data being substantially better than that of the egress lightspot. The latitudes of ingress and egress vary significantly ($-9^{\circ}$ versus $31^{\circ}$). It seems that the data better constrain zonal winds (and accordingly, radial distortions) at the ingress latitude. We also note that assuming longitudinal symmetry excludes possible localized events that could be happening in the atmosphere regions probed by ingress and egress, such as planetary-scale waves. When the ingress lightspot and egress lightspot are considered individually, different starting atmospheric profiles match best - the best match for the egress lightspot favors lower peak wind speeds, while the ingress lightspot data was best matched by the ALMA starting atmosphere and several doubled GCM zonal wind profiles (i.e., no favoring of low wind speeds as is observed for the egress lightspot). The strength of the match between the ingress lightspot observations and the atmospheric model derived from ALMA interferometry supports the complementarity of (optical) occultation and (radio) Doppler mapping techniques for zonal wind investigations.

Finally, the mid-event frame shows wider arcs on both the north and south limbs in a pattern that we were unable to reproduce with any of our forward models. We are looking forward to enriching the results of this paper with a refined inversion technique, analyzing a second occultation dataset gathered in 2022, and gathering additional data from the upcoming 2026 South America occultation.

Separately, the impending autumnal equinox suggests a critical time of rapid change for the zonal wind structure of Titan's atmosphere. We are fortunate to have one additional Titan stellar occultation of a bright ($G_{mag} = 10.62$) star approaching in September 2026. The center of the shadowpath will cross over southern Argentina and Chile, and we will employ a ``picket fence" line of portable telescopes, with chords crossing near the central flash of the occultation. The Very Large Telescope (VLT) is also within the shadowpath, and we hope to leverage the combination of the VLT and the portable telescopes to gain a comprehensive profile of Titan in the time immediately following northern autumnal equinox.

\begin{acknowledgements}
We would like to thank the staff astronomers and telescope operators at W. M. Keck Observatory for assisting with the operations both leading up to the observation night and during the event itself. TCM, EFY, and MAC were supported by a grant from the NASA ROSES Solar System Observations program. MAC and CAN were supported by a grant from the NASA ROSES Solar System Observations program.
\end{acknowledgements}

\bibliography{midatm}{}
\bibliographystyle{aasjournal}



\textbf{Appendix}

\begin{table}[ht]
\caption{Sum of squared errors table for model-data fits.\label{tab:si_table1}}

\begin{tabular}{||c | c | c | c||} 
 \hline
 Input atmosphere & Combined SSE & Ingress SSE  & Egress SSE\\ 
 \hline\hline
 $L_s 0^o$ & 882.0 & 57.9 & 824.1 \\ 
 \hline
 $L_s 10^o$ & 868.6 & 52.8 & 815.8 \\ 
 \hline
  $L_s 20^o$ & 885.2 & 61.9 & 823.4 \\ 
 \hline
  $L_s 30^o$ & 772.9 & 100.3 & 672.6 \\ 
 \hline
  $L_s 40^o$ & 766.6 & 109.4 & 657.2 \\ 
 \hline
  $L_s 50^o$ & 722.2 & 99.5 & 622.6 \\ 
 \hline
  $L_s 60^o$ & 720.2 & 68.7 & 651.5 \\ 
 \hline
  $L_s 70^o$ & 658.5 & 56.5 & 601.9 \\
 \hline
  $L_s 80^o$ & 623.6 & 58.4 & 565.2 \\ 
 \hline
  $L_s 90^o$ & 594.0 & 64.6 & 529.4 \\ 
 \hline
  $L_s 100^o$ & 584.1 & 71.5 & 512.6 \\ 
 \hline
  $L_s 110^o$ & 573.2 & 73.2 & 499.9 \\ 
 \hline
  $L_s 120^o$ & 570.5 & 73.8 & 496.7 \\ 
 \hline
  $L_s 130^o$ & 576.7 & 72.3 & 504.3 \\ 
 \hline
  $L_s 140^o$ & 587.7 & 67.1 & 520.6 \\ 
 \hline
  $L_s 150^o$ & 606.3 & 61.2 & 545.2 \\ 
 \hline
  $L_s 160^o$ & 651.6 & 57.3 & 594.3 \\ 
 \hline
  $L_s 170^o$ & 717.3 & 67.3 & 650.0 \\
 \hline
  $L_s 180^o$ & 807.7 & 67.3 & 740.4 \\
 \hline
  $L_s 190^o$ & 820.5 & 53.8 & 766.7 \\
 \hline
  $L_s 200^o$ & 881.4 & 74.2 & 807.2 \\
 \hline
  $L_s 210^o$ & 841.6 & 68.6 & 773.0 \\
 \hline
  $L_s 220^o$ & 932.2 & 71.0 & 861.2 \\
 \hline
  $L_s 230^o$ & 950.0 & 78.6 & 871.4 \\
 \hline
  $L_s 240^o$ & 966.6 & 72.5 & 894.1 \\
 \hline
  $L_s 250^o$ & 887.1 & 78.2 & 808.9 \\ 
 \hline
  $L_s 260^o$ & 803.5 & 90.3 & 713.1 \\
 \hline
  $L_s 270^o$ & 727.4 & 96.4 & 631.0 \\
 \hline
  $L_s 280^o$ & 678.7 & 98.2 & 580.5 \\
 \hline
  $L_s 290^o$ & 655.7 & 98.7 & 557.0 \\ 
 \hline
  $L_s 300^o$ & 651.7 & 96.0 & 555.6 \\
 \hline
  $L_s 310^o$ & 649.9 & 94.4 & 555.5 \\
 \hline
  $L_s 320^o$ & 672.8 & 92.4 & 580.4 \\
 \hline
  $L_s 330^o$ & 702.5 & 90.3 & 612.2 \\ 
 \hline
  $L_s 340^o$ & 780.4 & 84.9 & 695.5 \\
 \hline
  $L_s 350^o$& 861.9 & 68.5 & 793.4 \\
 \hline
  ALMA & 664.3 & 42.6 & 621.7 \\
\hline
\end{tabular} \\
\footnotesize{Sum of squared errors table of all model-data fits for both ingress and egress spots combined, ingress spots alone, and egress spots alone (36 GCM snapshots and 1 ALMA profile). }
\end{table}

\begin{table}
\caption{Sum of squared errors table for model-data fits (model amplitudes are scaled 2$\times$).\label{tab:si_table2}}

\begin{tabular}{||c | c | c | c||} 
 \hline
 Input atmosphere & Combined SSE & Ingress SSE  & Egress SSE\\ 
 \hline\hline
  $2\times L_s 0^o$ & 1263.6 & 160.1 & 1103.5 \\ 
  \hline
  $2\times L_s 10^o$ & 1633.3 & 276.1 & 1357.2 \\ 
  \hline
  $2\times L_s 20^o$ & 1625.4 & 585.8 & 1039.5 \\ 
 \hline
  $2\times L_s 30^o$ & 2237.3 & 1349.2 & 888.1 \\ 
 \hline
  $2\times L_s 40^o$ & 2175.5 & 1377.7 & 797.8 \\ 
 \hline
  $2\times L_s 50^o$ & 2062.2 & 1308.8 & 753.4 \\ 
 \hline
  $2\times L_s 60^o$ & 2023.8 & 1340.6 & 683.2 \\ 
 \hline
  $2\times L_s 70^o$ & 1811.2 & 1126.9 & 684.3 \\
 \hline
  $2\times L_s 80^o$ & 1769.7 & 843.1 & 926.6 \\ 
 \hline
  $2\times L_s 90^o$ & 1513.1 & 719.1 & 794.0 \\ 
 \hline
  $2\times L_s 100^o$ & 1277.4 & 562.0 & 715.5 \\ 
 \hline
  $2\times L_s 110^o$ & 1063.6 & 370.4 & 693.2 \\ 
 \hline
  $2\times L_s 120^o$ & 997.5 & 330.7 & 666.8 \\ 
 \hline
  $2\times L_s 130^o$ & 1103.6 & 406.3 & 697.3 \\ 
 \hline
  $2\times L_s 140^o$ & 1210.8 & 508.5 & 702.3 \\ 
 \hline
  $2\times L_s 150^o$ & 1529.7 & 746.2 & 783.5 \\ 
 \hline
  $2\times L_s 160^o$ & 1849.1 & 1083.5 & 765.6 \\ 
 \hline
  $2\times L_s 170^o$ & 1977.7 & 1304.0 & 673.7 \\
 \hline
  $2\times L_s 180^o$ & 2016.6 & 1246.7 & 770.0 \\
 \hline
  $2\times L_s 190^o$ & 2141.3 & 1040.3 & 1101.0 \\
 \hline
  $2\times L_s 200^o$ & 1346.2 & 272.9 & 1073.3 \\
 \hline
  $2\times L_s 210^o$ & 1406.0 & 219.9 & 1186.0 \\
 \hline
  $2\times L_s 220^o$ & 1383.8 & 92.3 & 1291.5 \\
 \hline
  $2\times L_s 230^o$ & 1382.8 & 36.6 & 1346.1 \\
 \hline
  $2\times L_s 240^o$ & 1349.7 & 28.9 & 1320.8 \\
 \hline
  $2\times L_s 250^o$ & 1406.6 & 30.8 & 1375.8 \\ 
 \hline
  $2\times L_s 260^o$ & 1740.8 & 38.8 & 1702.0 \\
 \hline
  $2\times L_s 270^o$ & 1376.4 & 26.0 & 1350.5 \\
 \hline
  $2\times L_s 280^o$ & 1219.1 & 45.3 & 1173.8 \\
 \hline
  $2\times L_s 290^o$ & 1171.4 & 52.1 & 1119.3 \\ 
 \hline
  $2\times L_s 300^o$ & 1228.4 & 46.9 & 1181.5 \\
 \hline
  $2\times L_s 310^o$ & 1247.0 & 40.0 & 1207.0 \\
 \hline
  $2\times L_s 320^o$ & 1200.5 & 44.1 & 1156.4 \\
 \hline
  $2\times L_s 330^o$ & 1237.0 & 38.2 & 1198.8 \\ 
 \hline
  $2\times L_s 340^o$ & 1434.0 & 30.3 & 1403.7 \\
 \hline
  $2\times L_s 350^o$& 1062.8 & 36.2 & 1026.5 \\
 \hline
  $2\times$ALMA & 1955.1 & 907.1 & 1048.1 \\
 \hline
\end{tabular} \\
\footnotesize{Sum of squared errors table of all model-data fits for both ingress and egress spots combined, ingress spots alone, and egress spots alone. The profiles correspond to input profiles with doubled magnitudes (36 GCM snapshots and 1 ALMA profile). }

\end{table}

\begin{table}

\caption{Sum of squared errors table for model-data fits (model amplitudes are scaled 0.5$\times$).\label{tab:si_table3}}
\begin{tabular}{||c | c | c | c||} 
 \hline
 Input atmosphere & Combined SSE & Ingress SSE  & Egress SSE\\ 
\hline\hline
 $0.5\times L_s 0^o$ & 583.4 & 86.4 & 497.0 \\ 
 \hline
 $0.5\times L_s 10^o$ & 581.4 & 81.2 & 500.3 \\ 
 \hline
  $0.5\times L_s 20^o$ & 569.0 & 75.5 & 493.5 \\ 
 \hline
  $0.5\times L_s 30^o$ & 565.0 & 74.6 & 490.3 \\ 
 \hline
  $0.5\times L_s 40^o$ & 553.2 & 70.9 & 482.3 \\ 
 \hline
  $0.5\times L_s 50^o$ & 545.8 & 75.6 & 470.1 \\ 
 \hline
  $0.5\times L_s 60^o$ & 540.6 & 74.1 & 466.5 \\ 
 \hline
  $0.5\times L_s 70^o$ & 534.2 & 79.1 & 455.0 \\
 \hline
  $0.5\times L_s 80^o$ & 516.4 & 80.0 & 436.3 \\ 
 \hline
  $0.5\times L_s 90^o$ & 522.2 & 82.6 & 439.5 \\ 
 \hline
  $0.5\times L_s 100^o$ & 524.9 & 83.2 & 441.8 \\ 
 \hline
  $0.5\times L_s 110^o$ & 512.8 & 84.7 & 428.1 \\ 
 \hline
  $0.5\times L_s 120^o$ & 518.0 & 84.4 & 433.6 \\ 
 \hline
  $0.5\times L_s 130^o$ & 514.4 & 83.9 & 430.5 \\ 
 \hline
  $0.5\times L_s 140^o$ & 519.9 & 80.2 & 439.7 \\ 
 \hline
  $0.5\times L_s 150^o$ & 528.0 & 81.7 & 446.3 \\ 
 \hline
  $0.5\times L_s 160^o$ & 529.2 & 78.7 & 450.5 \\ 
 \hline
  $0.5\times L_s 170^o$ & 542.2 & 79.8 & 462.4 \\
 \hline
  $0.5\times L_s 180^o$ & 556.6 & 75.2 & 481.5 \\
 \hline
  $0.5\times L_s 190^o$ & 563.1 & 76.6 & 486.6 \\
 \hline
  $0.5\times L_s 200^o$ & 591.0 & 78.6 & 512.4 \\
 \hline
  $0.5\times L_s 210^o$ & 601.8 & 81.5 & 520.3 \\
 \hline
  $0.5\times L_s 220^o$ & 612.9 & 84.1 & 528.8 \\
 \hline
  $0.5\times L_s 230^o$ & 607.8 & 86.3 & 521.5 \\
 \hline
  $0.5\times L_s 240^o$ & 598.8 & 84.4 & 514.4 \\
 \hline
  $0.5\times L_s 250^o$ & 583.0 & 85.9 & 497.1 \\ 
 \hline
  $0.5\times L_s 260^o$ & 557.3 & 87.2 & 470.1 \\
 \hline
  $0.5\times L_s 270^o$ & 557.2 & 89.8 & 467.4 \\
 \hline
  $0.5\times L_s 280^o$ & 542.9 & 88.3 & 454.6 \\
 \hline
  $0.5\times L_s 290^o$ & 543.8 & 90.5 & 453.3 \\ 
 \hline
  $0.5\times L_s 300^o$ & 529.8 & 90.3 & 439.5 \\
 \hline
  $0.5\times L_s 310^o$ & 534.7 & 90.3 & 444.4 \\
 \hline
  $0.5\times L_s 320^o$ & 538.0 & 87.5 & 450.4 \\
 \hline
  $0.5\times L_s 330^o$ & 546.2 & 85.9 & 460.3 \\ 
 \hline
  $0.5\times L_s 340^o$ & 557.8 & 87.1 & 470.7 \\
  \hline
  $0.5\times L_s 350^o$ & 579.0 & 87.8 & 491.3 \\
  \hline
  $0.5\times$ALMA & 524.5 & 72.9 & 451.5 \\
 \hline
\end{tabular} \\
\footnotesize{Sum of squared errors table of all model-data fits for both ingress and egress spots combined, ingress spots alone, and egress spots alone. The profiles correspond to input profiles with halved magnitudes (36 GCM snapshots and 1 ALMA profile). }

\end{table}


\end{document}